\pdfoutput=1
\documentclass[twocolumn, twoside,prl]{revtex4}
\usepackage{graphicx}

\usepackage{epsfig}

\usepackage{color}

\newlength {\wdig}
\newlength{\mstruth}
\newlength{\mstrutd}

\newcommand{\bq}[1]{\begin{equation}\label{#1}}
\newcommand{\eq}{\end{equation}}
\newcommand{\bqx}{\begin{equation}}
\newcommand{\eqx}{\end{equation}}
\newcommand{\bqa}{\begin{eqnarray}}
\newcommand{\eqa}{\end{eqnarray}}
\newcommand{\bqb}{\begin{eqnarray*}}
\newcommand{\eqb}{\end{eqnarray*}}

\newcommand{\bc}{\begin{center}}
\newcommand{\ec}{\end{center}}

\usepackage{latexsym}
\usepackage{graphicx}
\newcommand{\beq}{\begin{equation}}
\newcommand{\eeq}{\end{equation}}
\newcommand{\barr}{\begin{eqnarray}}
\newcommand{\earr}{\end{eqnarray}}

\newcommand{\ket}[1]{| #1 \rangle }
\newcommand{\bra}[1]{\langle #1 | }
\newcommand{ \ave}[2]{ \langle #1 | #2 | #1 \rangle }
\newcommand{\amp }[2]{\langle #1|#2 \rangle }

\newcommand{\upn}{^{(N)}}

\input colornam.sty

\begin{document}

\title{Quantum interference experiments, modular variables and weak measurements}

\author{Jeff Tollaksen$^{1}$}

\author{Yakir Aharonov$^{1,2}$}

\author{Aharon Casher$^{1,2}$}
\author{Tirzah Kaufherr $^{2}$}
\author{Shmuel Nussinov$^{1,2}$}
\affiliation{$^1$ Chapman University, Schmid College of Science, Department of Physics, Computational Science, and Engineering, 1 University
Dr, Orange, CA 92866, USA}

\affiliation{$^2$ School of Physics and Astronomy, Tel Aviv University,  %
Tel Aviv, Israel}

\begin{abstract}

We address the problem of interference using the Heisenberg picture and highlight some new aspects through the use of pre-selection, post-selection, weak measurements, and modular variables. We present a physical explanation for the different behaviors of a single particle when the distant slit is open or closed: instead of having a quantum wave that passes through all slits, we have a localized particle with non-local interactions with the other slit(s). We introduce a Gedanken-experiment to measure this non-local exchange. While the Heisenberg picture and the Schrodinger pictures are equivalent formulations of quantum mechanics, nevertheless, the results discussed here support a new approach to quantum mechanics which has led to new insights, new intuitions, new experiments, and even the possibility of new devices that were missed from the old perspective. 

\end{abstract}

\maketitle


\section{\bf I. Introduction}

\noindent The two-slit experiment 
is the quintessential example of the dual character of quantum mechanics. 
The initial incoming particle seems to behave as a wave when falling on the (left and right) slits, but when recorded on the screen, its wavefunction ``collapses" into that of a localized particle. 
By repeating the experiment for an ensemble of many particles, the interference pattern manifests through the density of hits along the screen  (aligned with, say the $x$ direction): $dn(x)/{dx}\sim |\psi_{\mathrm{L}}(x)+e^{i\alpha}\psi_{\mathrm{R}}(x)|^2$
with $\psi_{\mathrm{L}}(x)$ coming from the left slit, $\psi_{\mathrm{R}}(x)$ from the right (located a distance $D$ away), and $\alpha$ the relative phase between the left and right parts of the wavefunction 
(see fig. 1).  

\bigskip

\noindent There are two ways to think about such phenomenon:

\bigskip

{\bf The first} accepts the Schr\"{o}dinger description as given, with wavepackets evolving in time.  Indeed the Schr\"{o}dinger description has been extremely useful, 
having served, for example, as the starting point for the Feynman path integral.  
The apparent analogy between Schr\"{o}dinger wave interference and classical wave interference (arising from the use of identical calculations), presents a conceptually simple interpretation of quantum phenomena in terms of our classical picture.  One is often advised to apply this consistent formalism for statistical predictions (providing, in this case, probability distributions for the positions of {\em many} particles) {\em without} asking questions about it's interpretation. In fact, the 
belief that the Schr\"{o}dinger picture is the {\em only} way by which the interference and relative phase can be inferred, played a central role in the development of the probability amplitude interpretation in the quantum formalism.

\vskip-.1cm
\begin{figure}[tbph] 
  \centering
\includegraphics[width=\columnwidth]{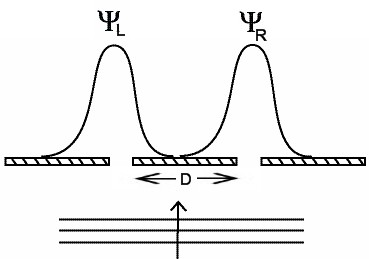}
\caption[]{\footnotesize Wavefunction for a single particle in double-slit setup when we do not know through which slit the particle has passed.}  
\label{mv1nleom}
\vskip-.3cm
\end{figure}

{\bf The second} way of thinking maintains that this is not the end of the story and advocates further inquiry.  For example, Feynman~\cite{feynman} stated that such phenomena ``..have in it the heart of quantum mechanics. In reality, it contains the only {\em mystery}."
Such proponents often seek to obtain as close a correspondence as possible between theory and measurement. As a consequence, they try to weed out ``classical" notions when they have been mis-applied to the quantum realm. For example, classical waves involve many degrees of freedom (e.g. field phenomenon such as sound and electromagnetic waves) and their phase {\em can} of course be measured by local experiments.
But the meaning of a quantum phase is very different.  
Multiplying the wavefunction $\psi_{\mathrm{L}}(x)+e^{i\alpha}\psi_{\mathrm{R}}(x)$ by an overall phase $\phi$ does not change the relative phase $\alpha$ and thus does not yield a different state.
Furthermore, it seems that the relative phase $\alpha$ cannot be measured directly on a single particle since it cannot be represented by a Hermitian operator.  That is, 
$\psi_{\mathrm{L}}(x)+e^{i\alpha}\psi_{\mathrm{R}}(x)$ and $\psi_{\mathrm{L}}(x)+e^{i\beta}\psi_{\mathrm{R}}(x)$
are not generally orthogonal and thus cannot be eigenstates belonging to different eigenvalues of a Hermitian operator.  
  In further contrast to the classical phase, a change in the relative quantum phase - say from $\psi_{\mathrm{L}}(x)+\psi_{\mathrm{R}}(x)$ to $\psi_{\mathrm{L}}(x)-\psi_{\mathrm{R}}(x)$ - would not result in a measureable change in any local properties.  The change only shows up in certain non-local properties or much later when the two separate components $\psi_{\mathrm{L}}(x)$ and $\psi_{\mathrm{R}}(x)$ eventually overlap and interfere.
It seems that the relative phase cannot be thought of simply as the difference between a local phase at $\psi_{\mathrm{L}}(x)$ and another local phase at $\psi_{\mathrm{R}}(x)$.

Another aspect of this second way of thinking is the realization that the Schr\"{o}dinger wave only has a measureable meaning for an ensemble of particles, not for a single particle.
This therefore leaves important questions unanswered concerning the \underline{\em physics} of interference from the perspective of a single particle: if physics obeys local dynamics, then how does the localized particle passing through the right slit sense whether or not the distant left slit is open (closed), causing it to scatter (or not scatter) into a region of destructive interference?  
Interference experiments have been performed with electron/photon beams whose intensity is sufficiently small such that only one electron/photon traverse the interference apparatus at a time.  The interference pattern with light and dark bands is nevertheless built up successively, mark by mark, with each individual ``particle-like" electron/photon,~\cite{feynman-hibbs}.  One is then confronted with the fact that a single degree of freedom created the interference pattern.  This mystery led Feynman to declare: ``Nobody knows how it can be like that." 
~\cite{feynman}

\bigskip

\noindent We follow the second way of thinking and offer a fresh approach to this time honored problem~\cite{danny}~\cite{app}~\cite{app2}~\cite{jtt}~\cite{atmv}.
To motivate the first step, involving a fundamental shift in the types of observables utilized, we make several observations: 
\begin{description}
\item{First}, most discussions of this problem are based on measurements which disturb the interfering particle.  This is one of the main reasons that quantum interference is generally considered to be intimately associated with the problems that stem from the statistical character of the quantal description.
\item{Second}, the observables studied to date have been simple functions of position and momentum.  These observables, however, are {\bf not} sensitive to the relative phase between different ``lumps" of the wavefunction (centered around each slit).  Nevertheless, the subsequent interference pattern of course is entirely determined by the relative phase between these ``lumps,"  suggesting that simple moments of position and momentum are not the most appropriate dynamical variables to describe quantum interference phenomena. 
\item {Third}, operators that {\bf \em are} sensitive to the relative phase are exponentials of the position and momentum.  
\end{description}

We address the first observation with non-disturbing measurements. To date, several non-disturbing measurements, such as weak measurements and protective measurements, 
have stimulated lively debates and have proven useful in separating various aspects of quantum theory from the probabilistic aspects~\cite{townes}.
The underlying framework for the approach to interference presented in this paper is based on another kind of non-disturbing measurement on the  ``set of deterministic operators" or ``deterministic experiments"~\cite{app}~\cite{atmv}~\cite{jtt}.  This set involves measurement of only those variables for which the state of the system under investigation is an eigenstate.  
This set answers ``what is the set of Hermitian operators $\hat{A}_\psi$ for which $\psi$ is an eigenstate?" 
for any state $\psi$, i.e.  $\hat{A}_\psi=\{ \hat{A}_i\,\, such \,\, that \,\, \hat{A}_i|\psi(t)\rangle = a_i|\psi(t)\rangle  \, , a_i\in \mathbf{\Re}\}$.
This question is dual to the more familiar question ``what are the eigenstates of a given operator?"  Measurement of these operators $\hat{A}_\psi$ does not collapse the wavefunction, since the wavefunction is initially  an eigenstate of the operator being measured.
Elaboration of this framework is left to existing and forthcoming literature~\cite{danny}~\cite{app}~\cite{app2}~\cite{jtt}. The essential point needed for this article is the relevance of deterministic experiments for a {\it single} particle since they can be performed without causing a disturbance.  

We address the second and third observations by performing yet another kind of non-disturbing measurement, namely weak measurements, on the observables 
that {\em are} sensitive to the relative phase.  These observables that are senstive to the relative phase are functions of {\bf\em modular variables}.   For the special case of interference in space, as considered here, the relevant modular variable is modular momentum, not ordinary momentum. These observables are also members of the ``deterministic set of operators" and are relevant for an individual particle. We then see that 
in the context of interference phenomenon, the Heisenberg equations of motion for these modular variables are {\em non-local}.
The nonlocality of these observables is quite intuitive: the operators sensitive to the relative phase simply translate the different ``lumps" of the wavefunction.  The appropriate translation may cause one lump to overlap with another lump or to overlap simply with the region where the distant slit is either open (or closed).   
This provides a {\bf physical} explanation for the different behavior of a {\em single} particle when the distant slit is open or closed.  It therefore provides the under-pinnings for a new ontology based on localized particles with non-local interactions, rather than an unphysical  Schr\"{o}dinger ``wave of probability" traveling throughout all of space.

This kind of non-locality which is revealed in the equations of motion, is {\bf \it dynamical} non-locality, to distinguish it from kinematic non-locality~\cite{shimony}~\cite{abepr} implicit in quantum correlations.  These two kinds of non-locality are fundamentally different:  kinematic non-locality arises from the structure of Hilbert space and does not create any change in probability distributions, causes and effects cannot be distinguished and therefore ``action-at-a-distance" cannot manifest.  Kinematic non-locality has been extremely useful, having catalyzed, e.g., much of the progress in quantum information science.~\cite{qis} 
 On the other hand, dynamical non-locality, arises from the structure of the equations of motion and {\em does} create explicit changes in probability, though in a ``causality-preserving" manner.  
This approach was first introduced by Aharonov, Pendelton and Petersen (APP)~\cite{app} in order to explain the nonlocality of topological phenomena such as the Aharonov-Bohm (AB) effect~\cite{ah1959}~\cite{abphystoday}.  
The AB effect conclusively proved that a magnetic (or electric) field inside a confined region can have a measureable impact on a charged particle which never traveled inside the region. 
In order to represent the closest correspondence between measurement and theory, APP introduced nonlocal interactions between the particle and field.  This was in contrast to the prevailing approach of reifying local interactions with (unphysical)  non-gauge invariant quantities outside the confined region, such as the vector (and/or scalar) potential.

Both dynamic and kinematic non-locality are generic and can be found in almost every type of quantum phenomenon.~\cite{popescu-1992} 
Prior to APP, dynamical nonlocality was avoided due to the possibility that it could  violate causality.  However, in a beautiful theorem, APP proved that the dynamical nonlocality they introduced could never violate causality.  They considered the general set of conditions necessary to see the non-local exchange of modular variables, for example when the left slit is either monitored or closed and the particle is localized around the right slit.  APP proved that these are precisely the same conditions which make the non-local exchange completely uncertain and therefore ``un-observable."

While it was beautiful that quantum mechanics allowed ``action-at-a-distance" to ``peacefully-coexist" with causality, this theorem nevertheless proved to be somewhat anti-climatic: if we cannot actually observe the nonlocal exchange of modular variable, then have we not violated the dictum of maintaining the closest correspondence between measurement and theory by claiming the existence of a new kind of nonlocal - yet un-observable - effect?

One of the principal new results 
presented in this paper is to show, for the first time, that these non-local interactions {\bf\em can} be observed.
This has to be done in a causality-preserving manner.  Therefore, in order to measure this nonlocality, we must utilize various tools such as 
pre-selection, post-selection, and weak measurements.  Although some of the components utilized in the present analysis were published long ago, they are not generally known and are  therefore briefly reviewed.

\bigskip

With this development, we have thereby underscored a  fundamental difference between classical mechanics and quantum mechanics that is easily missed from the perspective of the Schr\"{o}dinger picture:  the equations of motion for observables relevant to quantum mechanical interference phenomenon can be {\bf non-local} 
in a peculiar way that preserves causality.
These novel results motivate a new approach to quantum mechanics starting from the Heisenberg picture and involving the set of deterministic operators. 
While the new framework and associated language are, in principle, equivalent to the Schr\"{o}dinger formulation, it has led to new insights, new intuitions, new experiments, and even the possibility of new devices that were missed from the old perspective. These types of developments are signatures of a successful re-formulation.

Although further elaboration of this new approach is left to a future article~\cite{atmv}, we briefly mention one important conceptual shift:
when quantum mechanics is compared to classical mechanics, often 
the uncertainty or indeterminism of quantum mechanics is emphasized and the profound, fundamental differences in the dynamics is ignored.  This is perhaps a result of the similarity between the classical dynamical description (Poisson bracket) and the quantum dynamical description (commutator) for simple functions of momentum or position. Furthermore, uncertainty 
is viewed in a kind of ``negative" light: as a result of the uncertainty in quantum mechanicics, we have lost the ability that we had in classical mechanics to predict the future.
Not only is nature ``capricious," but it seems that we do not even gain anything from the uncertainty.

The new approach allows us to change this perspective by {\it deriving} uncertainty from principles that we argue are  more fundamental, namely from non-locality and causality.
This changes the meaning of uncertainty from one with a ``negative" connotation to one with a ``positive" connotation.
Something similar happened with special relativity when the
 axioms of relativity were discovered.  This inspired a modification of the old language: e.g. that light has the same
velocity in all reference frames is certainly highly unusual, but
everything works in a self consistent way due to the axiomatic framework, and because of this
special relativity is rather easy to understand. 
  
 Similarly, we are convinced that the new approach arising from this paper will lead to a deeper understanding of the nature of quantum mechanics.

\section{\bf II. Brief review of loss of interference from the Schr\"{o}dinger perspective}
We begin to motivate our approach by reviewing past attempts to analyze the disappearance of interference whenever it is possible to detect through which slit the particle passes.
The original debate was famously conducted by Einstein and Bohr.  Einstein attempted to challenge the consistency of quantum mechanics by arguing that a Which Way Measurement (WWM) could be performed without destroying the interference pattern by measuring the transverse recoil (i.e. the transverse momentum kick) of the double-slit screen after the particle passed through.  Bohr maintained that the consistency of quantum mechanics depended on the destruction of the interference pattern when WWM information is obtained.  He showed that the measurement-induced uncertainty created in the transverse position of the screen by an accurate  measurement of the transverse momentum was sufficient to destroy the interference pattern. 

This reasoning leads to a paradox which helps to motivate our approach. It has been argued (borrowing from  the discussion of the ``Heisenberg microscope") 
 that if the particle were ``observed" at the right slit,  then the photon involved in  this observation should have a wavelength ${\lambda}{\leq} D/2$ and a corresponding momentum uncertainty ${\Delta}p > 2\hbar/D$.
This momentum uncertainty is imparted to the particle making its wave number $k=p/\hbar$ uncertain, thereby destroying the interference pattern.

 This argument is incorrect. 
To see this, assume that a sensitive detector, placed at the left slit, failed to detect any particle.  We then know that all particles passed through the right slit. The interference pattern will then be completely destroyed despite the fact that there was no interaction with the detector!~\cite{app}~\cite{scully} 
One might suppose that since the action of opening/closing the left slit never caused an interaction with the particle at the right slit, then nothing associated with the particle should change.  But, it was first pointed out by APP~\cite{app} that in this scenario when a WWM is performed without actually interacting with the interfering particle, then the probability distribution of the momenta {\em does} change, although none of the moments of the momenta change. 

To best resolve this paradox, we need to take a step back.  We note that the effect of a generic interaction or collision between any two quantum systems 
can be  characterized by a change in the probability
   distribution of the momentum  i.e. going from an initial probability distribution, $\rho_i(p)$, to a final distribution,
 $\rho_f(p)$. 
We can analyze this change in two ways\footnote{We consider momentum here, but our comments apply to any conserved quantity.}:
   \begin{enumerate}
      \item Look at moments such as $\langle p^n \rangle=\int \rho(p)p^ndp$ and calculate
         $\delta\langle p^n\rangle=\langle p^n\rangle_f-\langle p^n\rangle_i$, and thus ask how the
         interaction affected these averages.  This is the usual approach. 
      \item Or, we may look at the fourier transform of the probability distribution $\int
         \rho(p)e^{\frac{i}{\hbar}pD}dp$.  (We will later see that these functions, $\langle e^{\frac{i}{\hbar}pD}\rangle$, are precisely the observables that are sensitive to the relative phase.)  To analyze the effect of the interaction, we calculate $\langle e^{\frac{i}{\hbar}pD}\rangle_f-\langle
         e^{\frac{i}{\hbar}pD}\rangle_i$
and ask how the interaction affected these averages.
   \end{enumerate}

   In principle, one can discuss the effect of interactions using (1) or (2), since knowing (2)
   for all $D$ is equivalent to knowing (1) for all $n$. 

\subsection{\bf II.a Analyzing changes in probability distribution  using method 1: moments of the conserved quantity}
 
Scully~\cite{scully} et al and Storey~\cite{storey} et al further debated the issues introduced by APP, resulting in many hundreds of cited papers.

Scully et al were dissatisfied with Bohr's original response to Einstein.  They suggested that a microscopic pointer (i.e. a micro-maser) could be used in such a way that 
 the interference in a WWM is destroyed without imparting any momentum to the particle (just as we alluded to earlier in the discussion of the case in which a sensitive detector failed to find the particle at the left slit).

However, Storey (et al) countered this, stating that the momentum distribution does change when WWMs are made.  
They noted that having a plane wave with initial $\Delta x=\infty$ and $\Delta p=0$ impinge on the 2-slits projects the initial plane wave onto ``lumps"  which therefore have a significant $\Delta p$.

The principal components of both camps' arguments were previously put forward in APP, i.e. there is both a change in probability and no change in the moments.  But, can we actually observe the change in the probability of the momentum when the left slit is open or closed?
To determine whether the momentum is disturbed by the WWM, the momentum of the particle must be known before the WWM and after.  However, if an ideal measurement is made of the momentum before the WWM, then we have effectively measured the interference, rendering useless the subsequent WWM.

The techniques of weak measurement have proven very useful in scenarios like this requiring manifestation of two opposing situations, i.e. to have a ``have-your-cake-and-eat-it" solution.
Weak measurements have had a direct impact on the central ``mystery" alluded to by Feynman concerning indeterminism, namely the fact that the past does not completely determine the future.
This mystery was accentuated by an assumed ``time-asymmetry" within quantum mechanics, namely the assumption that measurements only have consequences {\bf after} they are performed, i.e. towards the future.  Nevertheless, a positive spin was placed on quantum mechanic's non-trivial relationship between initial and final conditions by Aharonov, Bergmann and Lebowitz (ABL)~\cite{abl} who showed that the new information obtained from future measurements was also relevant for the {\bf past} of quantum systems and not just the future.  This inspired ABL to re-formulate quantum mechanics in terms of {\em pre- and post-selected ensembles}.
The traditional paradigm for ensembles is to simply prepare systems in a particular state and thereafter subject them to a variety of experiments.  These are ``pre-selected-only-ensembles." 
For pre-{\bf and-post-selected}-ensembles, we add one more step, a subsequent measurement or post-selection.  
By collecting only a subset of the outcomes for this later measurement, we see that the ``pre-selected-only-ensemble"   can be divided into sub-ensembles according to the results of this subsequent ``post-selection-measurement."  
Because pre- and post-selected ensembles are the most refined quantum ensemble, they are of fundamental importance and have revealed novel aspects of quantum mechanics that were missed before, particularly the weak value which has been confirmed in numerous weak measurement experiments.
Weak values have led to quantitative progress on many questions in the foundations of physics~\cite{at} including interference, etc.~\cite{danny}
 field theory, in tunneling, in quantum information such as the quantum random walk, in foundational questions, in the discovery of new aspects of mathematics,
 such as Super-Fourier or super-oscillations.  It has also led to generalizations of quantum mechanics that were missed before.  

While it is standard lore that the wave
 and particle nature cannot manifest at the same time,  weak measurements on pre- and post-selected ensembles {\it can} provide information about  both the
 (pre-selected) interference pattern and about the (post-selected) direction of motion for each particle.  This aspect of weak measurements formed the basis for the  first application of weak measurements to study the change in momentum for WWM within the double-slit setup as presented by Wiseman~\cite{wiseman}.  This was followed by an experiment(Mir, Lundeen, Mitchell, Steinberg, Garretson and Wiseman~\cite{steinberg2007}). 
Besides clarifying the different definitions and different measurements (etc) used by both sides of the debate, Wiseman and Mir et al show that the momentum transfer can be observed for the spatial wavefunction used in the 2-slits (as opposed to momentum eigenstates) by using weak measurements.

They implemented the weak measurement with position shifts and polarization rotations in a large optical interferometer.
Plotting the conditional probability to obtain a particular momentum (given the appropriate post-selection) 
and integrating over all possible post-selections, they were able to verify both the Scully and Storey viewpoints.  With respect to Scully~\cite{scully}, they show that none of the moments of the momentum change.  With respect to Storey~\cite{storey}, they show that the momentum does extend beyond a certain width.

However, there are inherent limitations to any approach based on analyzing changes in the probability for momenta through changes in the moments.  For example, while momentum is of course conserved, 
 there is no definite connection between the probability of an individual momentum before and after an exchange between the interfering particle and the slit.
Furthermore, the analysis in terms of moments does not offer any intuition as to {\bf\em how} or {\bf\em why} the probability of momentum changes.

\subsection{\bf II.b Analyzing changes in probability distributions using  using method 2: fourier transform of the conserved quantity}
When compared to the first (traditional) approach based on the moments, the second approach focusing on the fourier transform of the probability distribution has many advantages, both mathematical and physical.  
In this section, we briefly review some of the mathematical advantages, leaving much of the physical advantages to the rest of the article.

The first ``moments" approach to interference 
derived from intuitions developed with wavefunctions consisting of just one ``lump." In these cases, the averages of $x$ (or of $p$) evolve according to {\em local} classical equations of motion. Also the 
uncertainties 
  {  $(\Delta x)^2\equiv{\overline{(\hat{x})}^2-\overline{\hat{x}}^2}$ and $(\Delta p)^2\equiv{\overline{(\hat{p})}^2-\overline{\hat{p}}^2}$, describing the spread in these variables, have properties similar to those of the spread of variables in a classical situation with unsharply defined initial conditions} and which evolve according to diffusion-like rules.

This drastically changes when we have two or more separate ``lumps" of the wavefunction.
Indeed, the wavefunction, after passing through the symmetric two-slits, consists of a superposition of two identical, but physically disjoint ``lumps," $\psi_{\mathrm{L}}$ and $\psi_{\mathrm{R}}$ (see fig. 1):
\vskip-.2cm
\begin{equation}
\ket{\Psi_{\alpha}}=\frac{1}{\sqrt{2}}\{\ket{\psi_{\mathrm{L}}} +e^{i\alpha}\ket{\psi_{\mathrm{R}}}\}
\label{eq1}
\end{equation}
 Collapsing it to just $ \psi_{\mathrm{R}}(x)\equiv\langle x|\psi_{\mathrm{R}}\rangle$ does {\em not} change $\Delta p$ nor the expectation values of any finite order polynomial in $p$, as none of these local operators have a non-vanishing matrix element between the disjoint ``lumps" of the wavefunction. 
In other words, measuring through which slit the particle passes does not have to increase the uncertainty in momentum.    Later in this article we will review another uncertainty relationship which is more relevant for this issue.

Up until now we have focused on the disappearance of interference upon WWM. 
But the other fundamental mystery highlighted by Feynman remains: namely, how does a particle localized at the right slit ``know" whether the left slit is open or closed?  The first approach based on moments tell us nothing about this mystery.
The decisive importance of the second ``fourier transform" approach for this mystery is best illustrated through a basic theorem which characterizes all interference phenomenon:
all {\bf moments} of both position and momentum are {\em independent} of the relative phase 
   parameter $\alpha$  (until the wavepackets overlap):

\bigskip

 \noindent  {\bf Theorem I: } Let 
$\Psi_\alpha=\psi_{\mathrm{L}}(x,t)+e^{i\alpha}\psi_{\mathrm{R}}(x,t)$ 
 such that there is no overlap of $\psi_{\mathrm{L}}(x,0)$
   and $\psi_{\mathrm{R}}(x,0)$.  If $n$ and $m$ are integers, then for all values of $t$, and choices of $\alpha\,,\,\beta$:
\vskip -.6cm
   \begin{equation}\label{9.6}
\,\,\,\,      \int[\Psi_\alpha^*(x,t)\Psi_\alpha(x,t)-\Psi_\beta^*(x,t)\Psi_\beta(x,t)]x^mp^ndx=0
\label{thminterference}
   \end{equation}

\noindent For the particular double-slit wavefunction, it is easy to see that if there is no overlap between $\psi_{\mathrm{L}}$ and $\psi_{\mathrm{R}}$ then nothing of the form $\int_{-\infty}^{\infty}\Psi^* x^m p^n \Psi dx$ will depend  on $\alpha$ for any value of $m$ and $n$.  Furthermore, expanding $\int\{\psi_{\mathrm{L}}+e^{-i\alpha}\psi_{\mathrm{R}} \}^*x^m p^n\{\psi_{\mathrm{L}}+e^{i\alpha}\psi_{\mathrm{R}} \}dx$, we see that only the cross terms, i.e. $\langle \psi_{\mathrm{L}}| x^m p^n |e^{i\alpha}\psi_{\mathrm{R}}\rangle$, have the {\it possibility} of depending on $\alpha$; but operators of the form  $x^m p^n$ cannot change the fact that $\psi_{\mathrm{R}}$ and $\psi_{\mathrm{L}}$ do not overlap.  When integrated, these terms vanish and are therefore insensitive to the relative phase.

\bigskip

This suggests that these dynamical variables (e.g. $\langle x\rangle$, $\langle p\rangle$, $\Delta x$, $\Delta p$) are not the most appropriate to describe quantum interference phenomena.
What observables, then, are sensitive to this interference information which appears to be stored in a subtle fashion?
To fully capture the physics of these scenarios with wavefunctions composed of multiple lumps, non-polynomial and {\em non-local} operators, connecting the disjoint parts 
are required. 
 For many, equi-distant slits, these are the discrete translation by $\pm D$, namely $\exp\{\pm \frac{i}{h}\hat{p}D\}$, 
 effecting 
$\exp\{ -\frac{i}{h}\hat{p}D\}\psi_{\mathrm{R}}(x)\rightarrow\psi_{\mathrm{R}}(x-D)$ which overlaps with $\psi_{\mathrm{L}}(x)$.  The expectation value of
the translation operator $\exp\{ \frac{i}{h}\hat{p}D\}$ {\bf\em does} depend
on $\alpha$: 
$\langle \Psi_{\alpha}\mid\exp \{ i\hat{p}D/h\}\mid \Psi_{\alpha}\rangle=e^{-i\alpha}/2$.  

This provides the basis for a mechanism to explain {\bf\em how} the particle at the right ``knows'' what is happening at the left slit. As we will see, the second ``fourier transform" approach even provides us with the parameters relevant for this question (namely the distance between the slits), while the first ``moments" approach remains silent.  

 Before proceeding in the next section to the {\em physics} of interference for single particles, we briefly mention two additional mathematical advantages concerning the second ``fourier transform" approach.

First,  all the moments $\langle p^n\rangle$
   are averages of unbounded quantities, while $\langle  \exp\{ \frac{i}{h}\hat{p}D\}\rangle$
    are averages of bounded quantities. There are problems
   with unbounded quantities (as pointed out by Mir et al). Infinitesimal changes in $\rho(p)$ can cause very
   large changes in the moments $\langle p^n\rangle$. To see this, consider a negligible change, $\delta \rho(p)$, in $\rho(p)$.
   By negligible, we mean there is only a small change in the probability distribution. If we
   calculate $\delta\langle p^n \rangle=\int \delta \rho(p)p^ndp$, we could get a finite
   change if $\delta \rho(p)$ differs from zero at a sufficiently large $p$.
   In the limit, we could in fact consider $p\rightarrow\infty$ and $\delta
   \rho(p)\rightarrow 0$, in such a fashion that $\Delta p^n$ is finite.
   Then clearly $\delta\langle p^{n+1}\rangle$ diverges as do all higher moments. The second ``fourier transform" approach never has these kinds of problems and is always finite.

The other significant ``mathematical" difference concerns the utility of conservation laws.  
As mentioned in \S II.a,  while conservation of momenta is certainly maintained for the averages of moments, there is no definite connection between an individual momentum before and after an exchange in this general kind of setup.
As we shall see below, the second ``fourier transform" approach uncovers an exchange of a new conserved quantity.  The conservation law for these quantities can be expressed in a ``product-form" rather than a sum (as occurs for ordinary momentum).  This product-form conservation law is more relevant for many situations such as a change in relative phase.

\section{\bf III. Interference phenomenon from the Heisenberg perspective: modular variables}

As we have argued previously, the basic gauge symmetry would be violated if any quantum experiment could measure the local phase in $\ket{\Psi_{\alpha}}$ and therefore there is no {\em locally accessible phase} information in $\ket{\Psi_{\alpha}}$.
The relative phase is a truly {\em non-local} feature of quantum mechanics.  This point is often missed when the Schr\"{o}dinger picture is taught and classical intuitions are applied to interference.  For this and other reasons, we maintain that the non-local aspect of interference is clearer in the Heisenberg picture.

\subsection{\bf III.a Modular variables are the observables that are sensitive to the relative phase}
\label{sec4}

In \S II.b, we pointed to the significance of the Heisenberg translation operator, $\exp\{\pm \frac{i}{h}\hat{p}D\}$, 
 effecting  $\exp\{ -\frac{i}{h}\hat{p}D\}\psi_{\mathrm{R}}(x)\rightarrow\psi_{\mathrm{R}}(x-D)$ overlapping with $\psi_{\mathrm{L}}(x)$. Therefore, the expectation value of
the translation operator $\exp\{ \frac{i}{h}\hat{p}D\}$ {\bf\em does} depend
on $\alpha$: 
$\langle \Psi_{\alpha}\mid\exp \{ i\hat{p}D/h\}\mid \Psi_{\alpha}\rangle=e^{-i\alpha}/2$.  

\begin{figure}[tbph] 
  \centering
\includegraphics[width=\columnwidth]{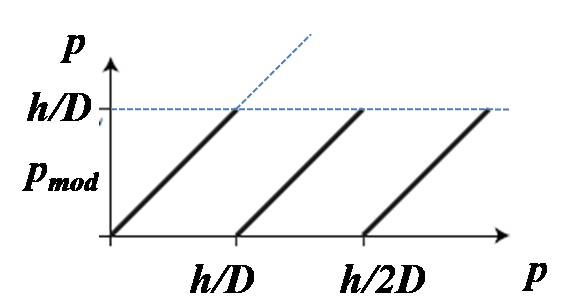}
\caption[]{\footnotesize Stacking  an integer number ($N_p$) of $\frac{h}{D}$ on top of the modular portion of $p$ ($p_{\mathrm{mod}}$.}  
\label{modmom}
\vskip-.5cm
\end{figure}

But, exactly {\em what} information about $\alpha$ does $\langle\exp\{\pm \frac{i}{h}\hat{p}D\}\rangle$ reveal?
It is easy to see that if we replace  $p$ with $p -\frac{nh}{D}$, then $e^{\frac{i}{\hbar}\hat{p}D}$ changes by $e^{\frac{iD}{\hbar}\frac{nh}{D}}=e^{in2\pi}=1$, i.e. nothing changes. 
Furthermore, suppose $n$ is the largest integer such that $n\frac{h}{D}<p$ (i.e. satisfying
$0\leq \hat{p} - n \frac{h}{D} \leq \frac{h}{D}$).  
This means that $e^{\frac{i}{\hbar}\hat{p}D}$ gives us information about the remainder after this integer number of $\frac{h}{D}$ is subtracted from $p$.  This is otherwise known of as the  modular momentum $p_{\mathrm{mod}}\equiv\hat{p}$ modulo $\frac{h}{D}$ (see fig. \ref{modmom}) defined by:
$\hat{p}$ modulo $\frac{h}{D} \equiv\hat{p} - n \frac{h}{D}$.

   It is clear that $p\,\, mod \,\, \frac{h}{D}$ has the topology of a
   circle, as would any periodic function.
Every point on the circle is another possible value for $p_{\mathrm{mod}}$.  We deal with modular quantities every time we look at a wristwatch which displays the time modulo 12.

We can get back to ordinary momentum through the relation:
\beq
p=N_p\frac{h}{D} + p_{\mathrm{mod}}
\eeq
We can see this (fig \ref{modmom}) if we stack  an integer number ($N_p$) of $\frac{h}{D}$ on top of the modular portion of $p$ ($p_{\mathrm{mod}}$ is the lower portion of fig \ref{modmom}).
Note that the eigenstates
of the translation operator $\exp\{ \frac{i}{h}\hat{p}D\}$ are
also eigenstates of the modular momentum $p_{\mathrm{mod}}$.

\vskip-.1cm

\subsection{\bf III.b For interference phenomenon, modular variables satisfy non-local equations of motion}
{ 
  
The key to our explanation of interference from the single particle perspective are the non-local equations of motion satisfied by these modular variables.  Thus, using $H=\frac{p^2}{2m}+V(x)$ and   
$e^{\frac{i}{\hbar}\hat{p}D}V(x)e^{-\frac{i}{\hbar}\hat{p}D}=V(x+D)$, 
we find {\bf non-local}~\cite{danny,app} Heisenberg equations of motion for modular variables:}
\begin{equation}
\frac{d}{dt}e^{\frac{i}{\hbar}\hat{p}D}=\frac{i}{\hbar}[H,e^{\frac{i}{\hbar}\hat{p}D}]
=\frac{i}{\hbar}[V(x)-V(x+D)]e^{\frac{i}{\hbar}\hat{p}D}
\label{nlham}
\end{equation}
with $e^{\frac{i}{\hbar}\hat{p}D}$ 
changing even when $\frac{\partial V}{\partial x}=0$.

This, essentially quantum phenomenon, has no classical counterpart.
The classical equations of motion for any function $f(p)$ derives from the Poisson bracket:
\beq
\frac{df(p)}{dt}=\left\{f(p),H\right\}_{PB}=-\frac{\partial f}{\partial p}{\,\,{\frac{\partial H}{\partial x}}\,\,\,\,\,\,}
+\underbrace{\frac{\partial f}{\partial x}}_{=0}\frac{\partial H}{\partial p}=0
\eeq
i.e. $f(p)$ changes only if $\frac{\partial V}{\partial x}\neq 0$
at the particle's location.

Unlike the Poisson bracket in classical mechanics, quantum mechanics has non-trivial and unique solutions to the commutator:  $[f(p),g(x)]=0$ if $f(p)=f(p+p_o)$, $g(x)=g(x+x_o)$ and $x_op_o=h$. These are easy to miss in the Schr\"{o}dinger picture.


However, in the Heisenberg picture, the non-local equations of motion 
that the modular variables satisfy  
show how the potential at the left slit {\em does} affect the evolution of the modular variable even when we consider a particle located at the right slit (and vice-versa, see fig. 3).  
{ Modular variables obey non-local equations of motion independent of the specific state of the Schr\"{o}dinger wavefunction, whether it is localized around one slit or in a superposition. Nevertheless, the modular momentum may change (non-locally) even if the wavefunction experiences no force.
We can therefore see that the non-local effect of the
open or closed slit is to produce a shift in the modular
momentum of the particle while leaving
the expectation values of moments of its
momentum unaltered.}

\subsection{\bf III.c Non-local exchange of modular variables in the double-slit setup}

For the special case of two-slits, a set of spin-like observables can be identified as members of the set of deterministic operators.  For simplicity (without affecting the  generality of our arguments),
 we can express the relevant modular variable as the parity (exchange) operation  $\hat{\textsf{P}}$ (effecting $\hat{\textsf{P}}\ket{\psi_{\mathrm{L}}}=\ket{\psi_{\mathrm{R}}}$  and  $\hat{\textsf{P}}\ket{\psi_{\mathrm{R}}}=\ket{\psi_{\mathrm{L}}}$).  It {\em \bf is} sensitive to the relative phase 
 $\alpha$ between the disjoint lumps of eq. \ref{eq1}~\cite{app}~\cite{danny}~\cite{jtt}: 
\begin{eqnarray}
\bra{\Psi_{\alpha}}\hat{\textsf{P}}\ket{\Psi_{\alpha}}\!\!&=&\!\! \frac{1}{{2}}\left\{\bra{\psi_{\mathrm{L}}} +e^{-i\alpha}\bra{\psi_{\mathrm{R}}}\right\}\hat{\textsf{P}}\left\{\ket{\psi_{\mathrm{L}}} +e^{i\alpha}\ket{\psi_{\mathrm{R}}}\right\} \nonumber\\
&=& \frac{1}{{2}}\left\{e^{i\alpha}+e^{-i\alpha}\right\}=  \langle \cos\alpha\rangle
\end{eqnarray}
To simplify further, we will focus on the $\pm 1$ eigenstates of $\hat{\textsf{P}}$:  
$\psi_{\mathrm{L}}(x)+\psi_{\mathrm{R}}(x)$ and $\psi_{\mathrm{L}}(x)-\psi_{\mathrm{R}}(x)$.
{ A measurement of which slit the particle goes through (i.e. a WWM) will change the value of $\langle\hat{\textsf{P}}\rangle$.  For example 
if the initial state is $\ket{\psi_{\mathrm{L}}} +\ket{\psi_{\mathrm{R}}}$, then $\langle\cos\alpha\rangle =1$, i.e. $\langle\hat{\textsf{P}}\rangle=1$.  If we collapse the state to $\ket{\psi_{\mathrm{R}}}$, then $\langle\cos\alpha\rangle =0$ and $\langle\hat{\textsf{P}}\rangle=0$.

\vskip-.1cm
\begin{figure}[tbph] 
  \centering
\includegraphics[width=\columnwidth]{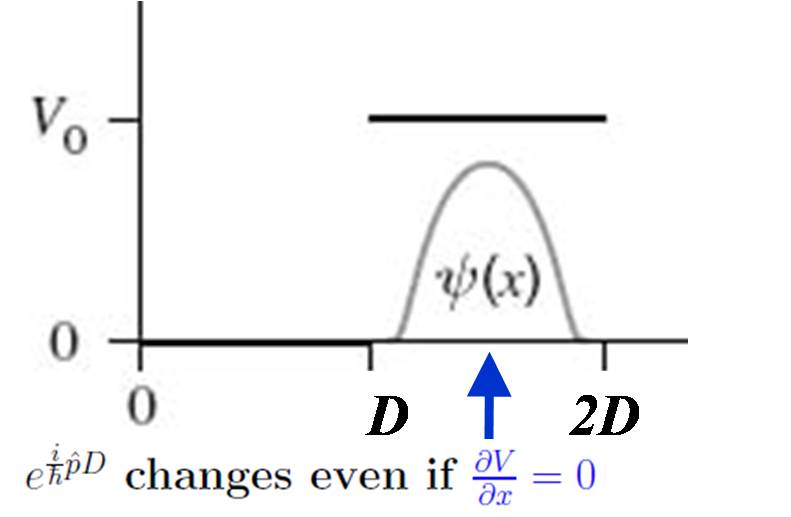}
\caption[]{\footnotesize A potential with 2 values and a wave-packet with support only in the interval $D <
x < DL$.}  
\label{modmomsteppot}
\vskip-.5cm
\end{figure}

We can also see from eq. \ref{nlham} that if the left slit is open, then $V(x)-V(x+D)=0$, and therefore $p_{\mathrm{mod}}$ is conserved.  However, if the left slit is closed, then $V(x)-V(x+D)\neq 0$ and $p_{\mathrm{mod}}$ is not conserved.

\subsection{\bf III.d Why does the interference pattern disappear when the particle is localized?}

When we obtain WWM information, we collapse the superposition from $\ket{\Psi_{\alpha}}$ to $\ket{\psi_{\mathrm{L}}}$ or $\ket{\psi_{\mathrm{R}}}$ (in the Schr\"{o}dinger picture).  In the Heisenberg picture, however, we cannot describe the collapse of a superposition.  The wavefunction is still of course relevant as a boundary condition, but it does not evolve in time. 
Only the operators evolve in time according to the Heisenberg equation of motion: 
$\frac{dA_H}{dt}=\frac{i}{\hbar}[H,A_H]+U^{-1}(t)\frac{\partial A_s}{\partial t} U(t)$.  But which operators become uncertain when WWM information is obtained?

Suppose again the particle travels through the right slit and we choose either to open or close the left slit.  This action causes a non-local exchange of modular momentum between the potential at the left slit and the particle going through the right slit.  Is this observable?

Up until this paper, it was believed that this could not be observabed.  The reason is that modular momentum (unlike ordinary momentum) becomes, upon detecting (or failing to detect) the particle at a particular slit, {\em maximally} uncertain.  
In other words, the effect of introducing a potential at a distance $D$ from the particle (i.e. of opening a slit) is equivalent to a rotation in the space of the modular variable - let's call it $\theta$ - that is exchanged nonlocally.   Suppose the amount of nonlocal exchange is given by $\delta\theta$ (i.e. $\theta\rightarrow\theta+\delta\theta$).  Now ``maximal uncertainty" means that the probability to find a given value of $\theta$ is independent of $\theta$, i.e. $P(\theta)=constant=\frac{1}{2\pi}$.  Under these circumstances, the shift in $\theta$ to $\theta+\delta\theta$ will introduce no observable effect, since the probability to measure a given value of $\theta$, say $\theta_1$, will be the same before and after the shift, $P(\theta_1)=P(\theta_1+\delta\theta_1)$.  We shall call a variable that satisfies this condition a ``completely uncertain variable".  Using this, APP proved a stronger {\it qualitative} uncertainty principle for the modular momentum, instead of the usual quantitative statement of the uncertainty principle (e.g. $\Delta pD\geq \hbar$): if the nonlocal exchange of any modular variable $\theta$ came close to violating causality, then the probability distribution for all averages of that modular variable flattens out, i.e  every value for $\theta$ became equally probable and change in $\theta$ 
becomes un-measureable:

\bigskip

 \noindent  {\bf Theorem II  qualitative uncertainty principle for modular variables}:  
 if $\langle e^{in\theta}\rangle=0$ for any integer $n\neq 0$ and  
if $\theta$ is a periodic function with period $\tau$, then $\theta$ is completely uncertain if $\theta$ is uniformly distributed on the unit circle. 
\label{proofcompleteunc}

\noindent Proof: we expand the probability density $Prob(\theta)$ to a
   Fourier series
      $Prob(\theta)=\sum_{n=-\infty}^{+\infty}a_ne^{in\theta}$ (integer $n$ is a requirement for the function to be periodic in $\theta$), where 
      $a_n=\int Prob(\theta)e^{in\theta}d\theta=\langle e^{in\theta}\rangle$
(since the average of any function is given the integral of the function with the probability).   We see that $Prob(\theta)=const$ if and only if $a_n=0$ for all $n\neq0$, and therefore $\langle e^{in\theta}\rangle=0$ for $n\neq0$.

\bigskip


Consider how this works in the double slit setup. Let us start with a particular $\ket{\Psi_{\alpha}}$, namely the symmetric 
($\alpha=0$, $\ket{\psi_{\mathrm{L}}} +\ket{\psi_{\mathrm{R}}}$) or anti-symmetric ($\alpha=\pi$, $\ket{\psi_{\mathrm{L}}} - \ket{\psi_{\mathrm{R}}}$) states.  The parity 
$\hat{\textsf{P}}$ then has sharp eigenvalues $\pm 1$.
However, $\hat{\textsf{P}}$ becomes maximally uncertain when the state is localized at one slit:
by definition, $\overline{\hat{\textsf{P}}^2}=1$, however, $\bra{\psi_{\mathrm{L}}}\hat{\textsf{P}}\ket{\psi_{\mathrm{L}}}\!\!=\!\!\bra{\psi_{\mathrm{L}}}{\psi_{\mathrm{R}}}\rangle\!\!=\!\!0$ so $\overline{\hat{\textsf{P}}}\!\!=\!\!0$, and therefore $\Delta \hat{\textsf{P}}\equiv\sqrt{\overline{(\hat{\textsf{P}})}^2-\overline{\hat{\textsf{P}}}^2}=\!\!1$, i.e. we have maximal uncertainty when the particle is localized at one slit.
Stated differently, when the particle is at the right (or left) slit its'
wave function is a superposition with {\em equal} weights of the two parity
eigenstates $\ket{\psi_{\mathrm{L}}} \pm\ket{\psi_{\mathrm{R}}}$ with $\pm 1$ eigenvalues which by
definition is the state of maximal variance of the operator involved.\footnote{
 In passing, we note that this is readily extended from the $Z(2)$ case of
just two slit to the $Z(N)$ case of $N$ equidistant, equal slits with periodic
boundary conditions (see Appendix B).}

The vanishing of the expectation value of the modular momentum variable is the manifestation in our present picture of the loss of information on $\alpha$ and of the interference pattern, once we localize the particle at the left or right slit.
\footnote{
Although much of the discussion in this article focuses on the simplest interference example with 2-slits, our approach becomes clearer when it is applied to an infinite number of slits with (interfering) particles that are initially in an eigenstate of momentum.  
In this case, we can directly speak about modular momentum (instead of slightly more complicated functions for the double-slit setup).  Also, 
both the non-local equation of motion for modular momentum is exact as is the conservation of modular momentum (see Appendix D).}

This brings us to what we believe to be a more physical answer (from the perspective of an individual particle) for the disappearance of interference: the momentum exchange with the left slit and resulting momentum uncertainty (destroying the interference pattern when the left slit is closed) is not that of ordinary momentum since as we noted $\Delta p$ does not change.  Rather, the closing of the left slit and localization of the particle at the right slit involves a non-local exchange of {\em modular} momentum.
\label{completeuncprin}
This phenomenon can also be demonstrated for any refinement of the double-slit.  For example, any measurement at the left slit introduces an uncertain potential there.  As a result of the non-local equations of motion, this introduces complete uncertainty in the modular variable.
Thus, detecting which slit a particle passes through  destroys all information about the modular momentum.

 It therefore appears that no observable effect of one slit acting on the particle traveling through the other slit can be obtained via the nonlocal equations of motion of the modular variable and therefore, this non-locality ``peacefully co-exists" with causality.
Have we not violated the dictum of maintaining the closest correspondence between measurement and theory by claiming the existence of a new kind of nonlocal - yet un-observable - effect?

The key novel observation we next make is that this non-locality {\em does have} an observable meaning in the context of weak 
measurements on pre- and post-selected ensembles~\cite{aav}~\cite{ar}

\section{\bf IV. Gedanken-experiment to measure non-local equations of motion}

\noindent What are the general issues involved in any attempt to measure this kind of nonlocality?  
\begin{description}
\item{\bf First:} if we start with the state $\psi_{\mathrm{R}}+\psi_{\mathrm{L}}$, i.e. a wavepacket around each slit, then the modular momentum is known but we cannot argue that the particle goes through one slit and is affected nonlocally by the other slit.  Therefore we need to start with a state which is localized around one slit.  
\item{\bf Second:} but under these circumstances when the particle is localized around one slit, the modular variable is completely uncertain and therefore un-observable.  How can we get around this fact in order to observe this nonlocality?  
\item {\bf Third:} if we are able to get around this fact, then how is causality not violated?
\end{description}
As we mentioned previously, weak measurements allows us to  ``have our cake and eat it" to a certain extent.  To address the first issue, we use pre- and post-selection to arrange for a localized particle property (pre-selection).  To address the second issue,  we later post-select a definite state of modular momentum. (We are interested in particular post-selections, rather than averages over all pre (and/or) post-selections as done in Mir et al~\cite{steinberg2007}.)  We may perform a weak measurement in order to see the weak value of the modular momentum. This weak measurement has a negligible probability to kick a particle centered around the right slit to the left slit, so we still satisfy the first criteria.  Finally, because we must rely on a post-selection and because of the nature of the weak measurement, it is impossible to violate causality with this method.

We proceed now to address each of these issues.

\subsection{\bf IV.a Information gain without disturbance: safety in numbers}

Traditionally, it was believed that if a measurement interaction is weakened so that there is no disturbance on the system, then no information will be obtained.   However, it has been shown that information can be obtained even though not a single particle (in an ensemble) was disturbed.  To set the stage, we consider a general theorem for any vector (state) in Hilbert space:
\bigskip

\noindent {\bf Theorem III:}
$\hat{A} |\psi \rangle = \langle\hat{A}\rangle \ket{\psi}  + \Delta A \ket{\psi_\perp}$
\label{thm1}

\bigskip

\noindent  where $\langle\hat{A}\rangle = \ave{\psi}{\hat{A}}$, $|\psi\rangle$ is any vector in Hilbert space, $\Delta A^2 = \ave{\psi}{(\hat{A} - \langle\hat{A}\rangle)^2}$, and $\ket{\psi_\perp}$ is a vector (state) in the perpendicular Hilbert space such that $\amp {\psi}{\psi_\perp} = 0$.

\noindent Proof:  left multiplication by $\bra{\psi}$ yields the first term; evaluating $|(A-\langle A\rangle)\ket{\psi}|^2 ={{\Delta}A}^2$ yields the second. 


So far, this is a completely general geometric property.  
To actually make a measurement of an observable $\hat{A}$, we switch on an interaction~\cite{vn} $H_{\mathrm{int}}={\lambda}g(t)\hat{q}\hat{A}$ with a normalized time profile $\int g(t)dt =1$. The pointer, namely the momentum $\hat{p}_{_{q}}$ conjugate to $\hat{q}$, shifts  by $\lambda \langle\hat{A}\rangle$. 

Now, the average of any operator $\langle\hat{A}\rangle\equiv\langle\Psi|\hat{A}|\Psi\rangle$ which appears in Theorem III, can be measured in three distinct ways~\cite{at3,spie-nswm}:  

\noindent {\bf 1. Statistical method with disturbance: } 
 the traditional approach is to perform ideal-measurements of $\hat{A}$ on each particle, obtaining a variety of different eigenvalues, and then manually calculate   
 the usual statistical average to obtain $\langle\hat{A}\rangle$.

\noindent  {\bf 2. Statistical method without disturbance: } 
The interaction $H_{\mathrm{int}}\!=\!-\lambda(t)\hat{q}\hat{A}$ is weakened 
 by minimizing $\lambda \Delta q$.  For simplicity, we consider $\lambda\ll 1$ (assuming without lack of generality that the state of the measuring device is a Gaussian with spreads $\Delta p_q\!=\!\Delta q\!=\!1$).   We may then set $e^{ -i \lambda  \hat{q} \hat{A}
 }\!\approx\! 1-i\lambda \hat{q} \hat{A}$ and use Theorem III
 to show that  the system state is:
   \begin{eqnarray}
e^{ -i  \lambda\hat{q} \hat{A}}|\Psi_{\mathrm{in}}\rangle&=&
(1\!-\!i\lambda \hat{q}\hat{A})|\Psi_{\mathrm{in}}\rangle\nonumber\\
&=&(1\!-i\!\lambda \hat{q}\langle\hat{A}\rangle)|\Psi_{\mathrm{in}}\rangle\!-i\!\lambda \hat{q}\Delta\hat{A}|\Psi_{\mathrm{in}\perp}\rangle
   \end{eqnarray}

\noindent Using the norm of this state ${\parallel (1-i\lambda \hat{q}\hat{A})|\Psi_{\mathrm{in}}\rangle
      \parallel}^2=1+{\lambda ^2\hat{q}^2}\langle \hat{A}^2\rangle$,
 the probability to leave $|\Psi_{\mathrm{in}}\rangle$ un-changed after the measurement is:
   \begin{equation}
      \frac{1+{\lambda ^2\hat{q}^2}{\langle\hat{A}\rangle}^2}
   {1+{\lambda ^2\hat{q}^2}\langle
   \hat{A}^2\rangle}\longrightarrow 1\,\,\,\,\,\,(\lambda \rightarrow 0)
   \end{equation}
while the probability to disturb the state (i.e. to obtain $|\Psi_{in\perp}\rangle$) is:
   \begin{equation}
      \frac{{\lambda ^2\hat{q}^2}{\Delta\hat{A}}^2}
   {1+{\lambda ^2\hat{q}^2}\langle
   \hat{A}^2\rangle}\longrightarrow 0\,\,\,\,\,\,(\lambda \rightarrow 0)
\label{collprob}
   \end{equation}
The final state of the measuring device is now a superposition of many substantially overlapping Gaussians with probability distribution given by $Pr(p_q)=\sum_i |\langle a_i|\Psi_{\mathrm{in}}\rangle|^2 \exp\left\{{-\frac{(p_q-\lambda a_i)^{2}} {2\Delta p_q^{2}}}\right\} $.  
This sum is a Gaussian mixture, so it can be approximated by a single Gaussian 
 $\tilde{\Phi}^{\mathrm{fin}}_{\mathrm{md}}(p_q)\approx\langle p_q|e^{-i\lambda \hat{q}\langle\hat{A}\rangle}|\Phi^{\mathrm{in}}_{\mathrm{md}}\rangle\approx\exp\left\{-{{(p_q-\lambda\langle\hat
 A\rangle)^2}\over{\Delta p_q^2}}\right\}$ centered on $\lambda\langle\hat{A}\rangle$.

\label{infogain}
It follows from 
eq.~\ref{collprob} that the probability for a collapse  decreases as $O(\lambda ^2)$, but the measuring device's shift grows linearly $O(\lambda)$, so $\delta p_q=\lambda a_i$~\cite{spie-nswm}.
For a sufficiently weak interaction (e.g. $\lambda\ll 1$), the probability for a collapse can be made arbitrarily small, while the measurement still yields information. However, the measurement  becomes less precise because 
the shift in the measuring device is much smaller than its uncertainty $\delta p_q\ll\Delta p_q$ (see fig \ref{seqm1}).

\noindent   {\bf 3. Non-statistical method without disturbance } is the case where $\langle\Psi|\hat{A}|\Psi\rangle$ is the ``eigenvalue" of a single 
``collective operator," $\hat{A}\upn\equiv \frac{1}{N} \sum_{\mathrm{i=1}}^{N} \hat{A}_i$ (with  $\hat{A}_i $ the same operator $\hat{A}$ acting on the $i$-th particle).
Using this, we are able to obtain information about $\langle\Psi|\hat{A}|\Psi\rangle$  without causing disturbance (or a collapse) and without using a statistical approach 
because any product state
  $|\Psi \upn\rangle$ becomes an eigenstate of the operator $\hat{A}\upn$. 
To see this, we apply Theorem III to the $N$ particle product state $\ket{\Psi\upn} =\ket{\psi}_1\ket{\psi}_2....\ket{\psi}_N$
 with all particles in the same state $\ket{\psi}$. We see that:  
\begin{equation}
\!\hat{A}\upn\ket{\Psi\upn}\!  =\! \frac{1}{N}\!\left[ N \langle\hat{A}\rangle\ket{\Psi\upn} + \Delta A \sum_i
|\Psi\upn_\perp(i) \rangle \right]
\label{avgop}
\end{equation}
where $\langle\hat{A}\rangle$ is the average for any one particle and the $N$ states $|\Psi\upn_\perp(i) \rangle = \ket{\psi}_1\ket{\psi}_2...\ket{\psi_\perp}_i...\ket{\psi}_N$ are mutually orthogonal. 
With a normalized state,  $|\Psi\upn_{\perp} \rangle = \sum_{i}\frac{1}{\sqrt{N}}|\Psi\upn_\perp(i) \rangle$, 
 the last term of eq. (\ref{avgop}) is $\frac{\Delta A}{\sqrt{N}}|\Psi\upn_{\perp} \rangle$ and $|\frac{\Delta A}{\sqrt{N}}
|\Psi\upn_{\perp} \rangle|^2\propto\frac{1}{N}$.
The probability that measuring $\hat{A}_i/N$ changes the state of the $i$-th system is proportional to $1/N^2$ and therefore the probability that it changes
the state of any system is proportional to $1/N$.  Thus, as $N \rightarrow \infty$, $|\Psi\upn\rangle$ becomes an eigenstate of $\hat{A}\upn$ with value
$\langle\hat{A}\rangle$
and not even a single particle has been disturbed (as $\hat{N} \rightarrow \infty$). 

To perform an actual measurement in this case, we fix $\Delta p_{q}$ (the width of the initial pointer momentum distribution) to be $1$.  
We can then take $\lambda\gg 1$,  allowing us to distinguish the result by having the shift, $\lambda\langle\hat{A}\rangle$, 
 exceed 
the width $\Delta p_{_{q}}=1$ of the distribution of the pointer.
In addition, fixing $\lambda\ll \sqrt{N}$ along with $|\hat{A}_i|<1$ ensures that the measurement does not shift any particle into an orthogonal state. 
The coupling to any individual member of the ensemble is reduced by $\frac{1}{N}$. When $N$ is very large, the coupling to individual systems is very weak, and in the limit $N\rightarrow\infty$, the coupling approaches zero. 
Although the probability that a measurement will disturb any member of the ensemble approaches zero as $\frac{1}{N}$, nevertheless, information about the average is obtained.

\subsection{\bf IV.b Pre-selection, post-selection and weak measurements}

By adding a post-selection to these ordinary -yet weakened- von Neumann measurements,  the measuring device will register  a weak value~\cite{aav}:
\begin{equation}
\hat{A}_w={{ {\langle \Psi _{\mathrm{fin}}\! \mid \hat{A} \mid \!\Psi _{\mathrm{in}} \rangle} \over {\langle \Psi _{\mathrm{fin}} \!\mid \!\Psi _{\mathrm{in}} \rangle}}}
\label{expweak}
\end{equation}
with $ \ket{\Psi_{\mathrm{in}}} $ and $ \ket{\Psi_{\mathrm{fin}}}$ the initial and final (post-selected) states.
The weak-value, $A_{\mathrm{w}}$, is an unusual quantity and is not in general an eigenvalue of $\hat{A}$.
 We have used such limited disturbance measurements to explore many  paradoxes (see, e.g. ~\cite{at,jtt,popescu}).  A number of experiments have been performed to test the predictions made by weak measurements and results have proven to be in very good agreement with
 theoretical predictions \cite{RSH,Ahnert,Pryde,Wiseman,Parks}.
\begin{figure}[tbph] 
  \centering
\includegraphics[width=\columnwidth]{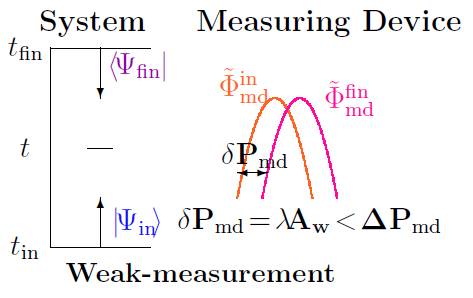}
\caption[]{\footnotesize if a weak-measurement is performed at $t$ (characterized e.g. by $\delta p_q\!=\!\lambda \!A_w\!\ll\!\Delta p_q$), then the outcome of the weak-measurement, the weak-value, can be calculated by propagating the
state \textcolor{RedViolet}{$\langle\!\Psi_{\mathrm{fin}}\!\!\mid$} backwards in time from $t_{\mathrm{fin}}$
to $t$ and the state \textcolor{BlueViolet}{$|\!\Psi_{\mathrm{in}}\!\rangle$} 
forwards in time from 
$t_{\mathrm{in}}$ to $t$; the weak-measurement does not cause a collapse and thus no new boundary condition is created at time $t$.}
\label{seqm1}
\end{figure}

Eq. \ref{expweak} can also be motivated by inserting a complete set of states $\{ \ket{\Psi_{\mathrm{fin}}}_j \}$ into $\langle\hat{A}\rangle$ 
\begin{equation}
\langle\hat{A}\rangle\!\!=\!\!\bra{\Psi _{\mathrm{in}}}\hat{A}\ket{\Psi _{\mathrm{in}}}\!\!=\!\!\sum_j |\langle \Psi _{\mathrm{fin}} \!\mid_j \!\Psi _{\mathrm{in}}
\rangle|^2\ 
\!\!\underbrace{{ {\langle \Psi _{\mathrm{fin}}\! \mid_j \hat{A} \mid \!\Psi _{\mathrm{in}}
\rangle} \over {\langle \Psi _{\mathrm{fin}} \!\mid_j \!\Psi _{\mathrm{in}}
\rangle}}}_{A_{\mathrm{w}}^j\equiv weak\,\,value}
\label{expweak2}
\end{equation}
with $ \ket{\Psi_{\mathrm{fin}}}_j $ the states corresponding to the outcome of a final ideal measurement on the system (i.e. the post-selection).
The average $\langle\hat{A}\rangle$ over all post-selections $j$ is thus constructed out of pre- and post-selected sub-ensembles in which the weak value ($A_{\mathrm{w}}^j$) is multiplied by a probability to obtain the particular  post-selection $\ket{\Psi_{\mathrm{fin}}}_j$.

To see more precisely how the weak value arises naturally from this weakened measurement with post-selection, we consider the final state of the measuring device after the above described procedure described in the third ``non-statistical" method:
\begin{eqnarray}
|\Phi_{\mathrm{fin}}^{\mathrm{MD}}\rangle\!&=&\!\prod_{j=1}^N \langle\Psi_{\mathrm{fin}}|_j \exp(\frac{-i\lambda}{N}  \hat{q} \sum_{\mathrm{k=1}}^N  \hat{A}_k)  \!\prod_{i=1}^N|\Psi_{\mathrm{in}} \rangle_i |\Phi_{\mathrm{in}}^{\mathrm{MD}}\rangle\,\,\,\,\nonumber\\
&=& \prod_{j=1}^N \langle\Psi_{\mathrm{fin}}|_j \exp(\frac{-i\lambda}{N}  \hat{q}\hat{A}_j)  |\Psi_{\mathrm{in}} \rangle_j |\Phi_{\mathrm{in}}^{\mathrm{MD}}\rangle
\label{eqexp}
\end{eqnarray}
Since the particles do not interact with each other,
we calculate one term and take the result to the $Nth$ power. 
(In the following, we substitute the parity operator, $\hat{\textsf{P}}$, for $\hat{A}$.) 
Using $\hat{\textsf{P}}^2=1$, eq. \ref{eqexp} becomes:
\begin{eqnarray}
&&\!\!\!\!\!\!\!\!\!\!\!\!\!\!\!\!\!\!\!\!\!\!\!\!|\Phi_{\mathrm{fin}}^{\mathrm{MD}}\rangle \!\!=\!\! \left\{\langle\Psi_{\mathrm{fin}}| (\cos \frac{{\lambda} \hat{q}}{N}-i\hat{\textsf{P}}\sin \frac{{\lambda} \hat{q}}{N})  |\Psi_{\mathrm{in}} \rangle\right\}^N \!\!\!\!|\Phi_{\mathrm{in}}^{\mathrm{MD}}\rangle\nonumber\\
\!\!\!\!&=& \!\! (\langle\Psi_{\mathrm{fin}}|\Psi_{\mathrm{in}} \rangle)^N 
\!\left\{\!\cos \frac{{\lambda} \hat{q}}{N}\!-\! i\textsf{P}_{\mathrm{w}}\sin \frac{{\lambda} \hat{q}}{N}\right\}^N \!\!\!\!\! |\Phi_{\mathrm{in}}^{\mathrm{MD}}\rangle
\label{bigspina}\\
&\approx &\!\!\!\!
\left\{\!1\!\! - \! i\textsf{P}_{\mathrm{w}} \!\frac{\lambda \hat{q}}{N}\!+\!\cdot\cdot\! \right\}^N\!\!\!\!|\Phi_{\mathrm{in}}^{\mathrm{MD}}\rangle\!\approx\!\exp(-i{\lambda}  \hat{q}\textsf{P}_{\mathrm{w}}\!)|\Phi_{\mathrm{in}}^{\mathrm{MD}}\rangle
\label{finwm}
\end{eqnarray}
The first bracket of eq. \ref{bigspina} can be neglected since it does not depend on $\hat{q}$ and thus can only affect the normalization.  
Eq. \ref{finwm} represents a shift in the pointer by the weak value, $\textsf{P}_{\mathrm{w}}$, i.e. $\Phi_{\mathrm{in}}^{\mathrm{MD}}(p_{_{q}})\rightarrow \Phi_{\mathrm{fin}}^{\mathrm{MD}}(p_{_{q}}-\lambda \textsf{P}_{\mathrm{w}})$.

\subsection{\bf IV.c Applying pre- and post-selection and weak measurements to interference phenomenon}

How can we use these tools to perform measurements of dynamical nonlocality?
 To briefly summarize our procedure, we start with particles sent through the right slit.  Before they encounter the double-slit, we perform a weak measurement of the modular momentum (which, again, {\em \bf is} sensitive to the relative phase).  We then choose whether to open the left slit or to close it.  After the particles pass the double-slit setup, we perform an ideal measurement of the modular momentum and post-select only those particles in a particular eigenstate of this modular momentum.
When we analyze the earlier weak measurement (assuming the post-selection is satisfied), we see two dramatically different results: one result if the left slit is  closed and a very different result if the left slit is opened.  The slit is open or closed only after the weak measurement has been completed and the results recorded. 

In this section we will use the third ``non-statistical" method and will later discuss the use of the second ``statistical" method.
Consider the following sequence: 
\begin{description}
\item {\bf a)} 
we send towards the slits, 
 $N$ consecutive particles,  
each in the same state centered around the right slit, ${\Psi_{\mathrm{R}}}$, i.e. we pre-select $\ket{\Psi_{\textrm{in}}}=\ket{\psi_{\mathrm{R}}}$ rather than $\ket{\Psi_{\alpha}}$;
\item {\bf  b)} after the pre-selection, but prior to encountering the slits, we measure weakly  the average modular variable: $\hat{\textsf{P}}\upn$, i.e. we  weakly measure the average modular variable (the parity) with an outcome of $\cos\alpha$.
In order to perform this measurement, we utilize (following von Neumann) the interaction Hamiltonian $H_{\mathrm{int}}=\frac{1}{N} \sum_{\mathrm{i=1}}^{N} \lambda g(t) \hat{q}\hat{\textsf{P}}_i  $ thereby generating the evolution $\Pi_{\mathrm{i=1}}^{N}\exp\{- i\lambda \hat{q} \hat{\textsf{P}}_i\}$ which simply sums the displacements of the ``pointer" 
due to the interactions with each of the N particles, namely a shift proportional to $\hat{\textsf{P}}\upn$;
\item {\bf  c)} finally, we post-select an eigenstate of the same modular variable observable previously measured weakly.  In particular, we post-select the symmetric state:
$\ket{\Psi_{\textrm{fin}}}=\frac{1}{\sqrt{2}}\{\ket{\psi_{\mathrm{L}}} +\ket{\psi_{\mathrm{R}}}\}$, which we note 
is an eigenstate of  the corresponding parity operator $\hat{\textsf{P}}$ with eigenvalue $+1$.
\end{description}
What then are the results of the above weak measurement,  
1) when both slits are open and 
 2) when the left slit
is closed:

\bigskip

{\bf \textsf{Case 1:} With the left slit open,} parity is conserved since in this symmetric slit arrangement, the Hamiltonian commutes with the parity operator.  (Furthermore, as noted in  \S III.b by eq. \ref{nlham}, $\frac{d}{dt}e^{\frac{i}{\hbar}\hat{p}D}=0$ when the left slit is open because $V(x)-V(x+D)=0$.)
This can also be seen by evolving the post-selected state  backward in time yields then $\hat{\textsf{P}}_i=+1$ for each of the $N$ particles (both before and after the double-slit)
and the measuring device then registers the weak value: 
 ${\textsf{P}}\upn_{w}=1$.  More specifically, the wavefunction of the measuring device evolves as:
\begin{eqnarray}
\Phi_{\mathrm{fin}}^{\mathrm{MD}}(p_{_{q}}) \!\!& \approx & \!\!
\left\{\!\left\{\!\bra{\psi_{\mathrm{L}}}\!+\!\bra{\psi_{\mathrm{R}}} \right\} \! \exp\!\left\{\!-i\frac{\lambda}{N} \hat{q}{\textsf{P}_{\mathrm{w}}}\! \right\} \!\ket{\psi_{\mathrm{R}}}\!\right\}^N\!\Phi_{\mathrm{in}}^{\mathrm{MD}}(p_{_{q}})\nonumber\\
& = & \lbrack\exp\left\{-i\frac{\lambda}{N} \hat{q}\right\}\rbrack^N \Phi_{\mathrm{in}}^{\mathrm{MD}}(p_{_{q}})= \Phi_{\mathrm{in}}^{\mathrm{MD}}(p_{_{q}}-\lambda)
\end{eqnarray}

{\bf \textsf{Case 2:} With the left slit closed,} the results of the weak measurement described above are {\em drastically} changed. Parity is now maximally violated and there is no connection between the $+1$ parity of the post-selected state  and the results of the weak parity measurements 
performed prior to entering the slits. (Again, as noted by eq. \ref{nlham} in \S III.b, $\frac{d}{dt}e^{\frac{i}{\hbar}\hat{p}D}\neq 0$ when the left slit is closed because $V(x)-V(x+D)\neq 0$.)

We next show that with this second ``slit-closed" case, the weak value of the parity is centered around $0$. Only $\ket{\psi_{\mathrm{R}}}$ can now propagate through the system of slits (any component of $\ket{\psi_{\mathrm{L}}}$ generated by the weak measurement is  always reflected by the closed slit.\cite{omit3}) 
The pointer shift 
is given by  eq. \ref{bigspina}
\begin{equation}
\Phi_{\mathrm{fin}}^{\mathrm{MD}}(p_{_{q}})\!=\!\left\{\!\!\bra{\psi_{\mathrm{R}}}\cos \frac{\lambda \hat{q}}{N} - i\hat{\textsf{P}}\sin\frac{\lambda \hat{q}}{N}\ket{\psi_{\mathrm{R}}}\!\!\right\} ^N \!\!\!\!\Phi_{\mathrm{in}}^{\mathrm{MD}}(p_{_{q}})
\label{eq15}
\end{equation}
Using $\bra{\psi_{\mathrm{R}}}\hat{\textsf{P}}\ket{\psi_{\mathrm{R}}}=\bra{\psi_{\mathrm{R}}}\psi_{\mathrm{L}}\rangle=0$, only the cosine part remains.  The pointer state $\Phi_{\mathrm{in}}^{\mathrm{MD}}(p_{_{q}})$ then shifts by:
\begin{equation}
\Phi_{\mathrm{fin}}^{\mathrm{MD}}(p_{_{q}})=\left\{\frac{1}{2}\right\}^N\left\{e^{\frac{i\lambda }{N}\hat{q}}+e^{-\frac{i\lambda }{N}\hat{q}} \right\}^N\Phi_{\mathrm{in}}^{\mathrm{MD}}(p_{_{q}})
\end{equation}
which upon binomial expansion becomes
\begin{equation}
\Phi_{\mathrm{fin}}^{\mathrm{MD}}(p_{_{q}})=\left\{ \frac{1}{2}\right\}^N \sum_k { {N}\choose{k}} \Phi_{\mathrm{in}}^{\mathrm{MD}} (p_{_{q}}+\frac{\lambda(2k-N)}{N})
\end{equation}
Since the binomial coefficients ${{N}\choose{k}}$ peak around $k=\frac{N}{2}$, 
the effect of the shifts vanishes in the $N\rightarrow \infty$ limit, and 	$\langle{\textsf{P}}\upn\rangle_{\mathrm{w}}=0$ as claimed.

The {\it earlier} weak measurement of the parity yields $\langle\hat{\textsf{P}}\upn\rangle_w=0$ (case 2) if the left slit is ({\it later}) closed and 
$\langle\hat{\textsf{P}}\upn\rangle_w=1$ (case 1) if the left slit is ({\it later}) opened. 
All incoming particles are initially in the state $\ket{\psi_{\mathrm{R}}}$, so we would not expect that closing  the left slit should have any effect on the result of any weak measurement (and in particular weak measurements performed prior to the opening or closing of the slits).  

\section{\bf V. Discussion}

How do we understand these two results? 
In principle, a weak measurement with finite $N$ shifts particles from the right slit to left slit so that the evolving wave-packet has  $\ket{\psi_{\mathrm{L}}}$ components and therefore may ``sense"  whether the left slit is open or closed.  However, 
all modular operators and parity in particular 
have norms $\leq 1$.  The exponents in the von Neumann interaction Hamiltonian are thus bound by 
$\frac{\lambda }{N}\hat{\textsf{P}}\upn\hat{q}< \frac{\lambda}{N} \hat{q}$
and hence it suffices to expand the binomial to order of a few $\lambda$.  This implies then that the weak measurement {\bf can shift at most a few ($\lambda$) of the $N$ particles from the right slit to the left slit.}
But, how can 
the $N-\lambda$ particles which were not shifted, and did not go through the left slit still be influenced non-locally so that we will have the dramatic (and large) change from case 1 (each particle shifts the measuring device by $\langle\hat{\textsf{P}}\rangle_w=1$) to case 2 (each particle shifts the measuring device by $\langle\hat{\textsf{P}}\rangle_w=0$)?

We do not see any reasonable way to use local interactions at the left-hand slit to account for the different subsequent behavior of the particles going through the right slit.

We can, however, make sense of the results by considering the non-local behavior of modular variables.
In particular, the first result is calculated  by using the non-local exchange of modular momentum.  The second result is calculated by using {\em conservation} of modular momentum. The use of this conservation principle is one of the crucial features that distinguishes our procedure from any observation that could be done with ordinary momentum.

These issues are not just  ``academic," as this article sets the stage for a forthcoming paper~\cite{nlexp} describing an actual quantum optics experiment to measure the non-local exchange. 
For illustrative purposes, we mention an experimentally simpler example using the second method of \S IV.a, i.e. the statistical weak measurement, unlike the weak measurement of a collective observable pre-scribed in section IV.a. 
Consider two consecutive Mach-Zehnder interferometers (see fig. 4).  The first Mach-Zehnder prepares the pre-selection: by  adjusting the arm lengths, it  is possible to arrange that the photon emerges at $R_4$,  which corresponds to the pre-selection $\ket{\psi_{\mathrm{R}}}$ (localized at the right slit).\footnote{We label a left-pointing arm as  $|L_n\rangle$, where a subscript $1$ refers to the photon before entering $BS_1$; a $2$ refers to the photon after entering $BS_1$ and before $M_1$ (etc); similarly $|R_n\rangle$ for a right-pointing arm.   
When put into the MZI in the right arm (without any weak measurement), the photon will exclusively exit at $R_4$.  Specifically, $|R_1\rangle \stackrel{BS_1}{\Longrightarrow} \frac{1}{\sqrt{2}}\lbrace i|L_2\rangle + |R_2\rangle\rbrace\stackrel{M_1/M_2}{\Longrightarrow}\frac{1}{\sqrt{2}}\lbrace i|L_3\rangle - |R_3\rangle\rbrace
 \stackrel{BS_2}{\Longrightarrow}  \frac{1}{2}\lbrace i|L_4\rangle -|R_4\rangle\rbrace-\frac{1}{2}\lbrace |R_4\rangle +i|L_4\rangle\rbrace=-|R_4\rangle$.}
In addition,  the weak measurement of parity is performed within the first Mach-Zehnder
by measuring small transverse shifts in the position of the photon produced by inserting thin glass plates\cite{stein1}  on both the $R_2$ and $L_2$ arm.   
The regime of weak measurement is obtained by adjusting the tilt of the plates so that the transverse spatial shift is small compared to the uncertainty in the transverse position of the photon.
The second Mach-Zehnder is the analog of the double-slit: e.g. blocking the $L_4$ path corresponds to closing the left-slit of the double-slit setup. 


\begin{figure}[tbph] 
  \centering
\includegraphics[width=\columnwidth]{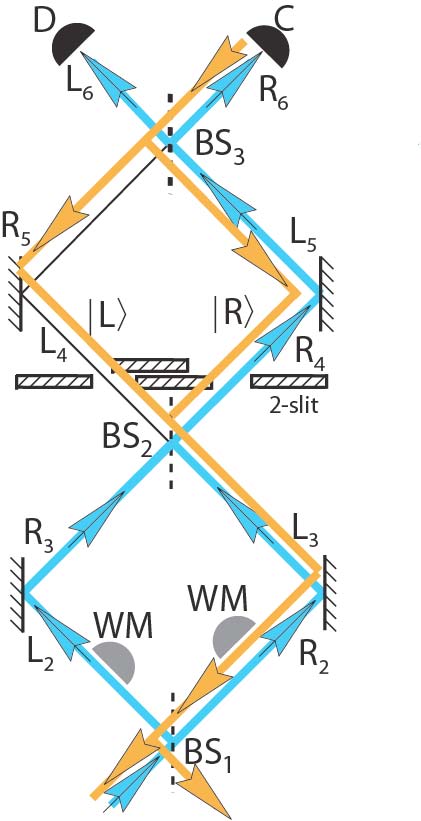}
\caption[]{\footnotesize The Mach-Zehnder analog of the 2-slit setup.  Also depicted is the  pre-selected wavefunction (evolving forward in time in blue) and the post-selected wavefunction (evolving backward in time - orange).}
\label{2mv}
\vskip.5cm
\end{figure}

Note: when the particle passed through the 2-slits (after $BS_2$), then parity was a nonlocal operator.  Later in time (after $BS_3$), the previously nonlocal parity was converted into a local quantity.  Because of this feature, we are able to perform an ideal measurement - i.e. a post-selection - of the parity. 

 If we post-select the $+1$ parity (after the photon passes $BS_3$) and if $L_4$ is open, then the earlier weak measurement of parity will register $+1$ (meaning that the weak value of the number of particles within arm $R_2$ is $(N_{R_{2})_\mathrm{w}}=1$ and within arm $L_2$ is $(N_{L_{2})_\mathrm{w}}=0$).  However, if we now close the left-hand slit, i.e. semi-arm $L_{4}$, then then the earlier weak measurement will register $(N_{R_{2})_\mathrm{w}}=(N_{L_{2})_\mathrm{w}}=\frac{1}{2}$  even though no particles took the path where the slit was closed!\footnote{Again, the relation with modular variables is somewhat clearer if we consider many slits since modular momentum is exactly conserved and we can directly speak about modular momentum rather than parity in the 2-slit case.  Suppose we send particles that are localized around a single slit towards the infinite slits and we then perform a weak measurement of $\sum \exp \{ \frac{i}{\hbar}\hat{p}D\}$.  We then consider 2 situations:  either all the other slits are open or all the other slits are closed.  In those cases in which the post-selection yields an eigenstate of modular position, and if all the slits are open, then we see that modular momentum is conserved and the earlier weak measurement will equal the post-selected ideal measurement.  If the other slits are closed, then modular momentum is not conserved, and the non-local equations of motion will reflect this.}

The third ``non-statistical" method mentioned in \S IV.a is also extremely important for many reasons.   For example, it can be used to measure a very wide variety of 
Hermitian observables involving highly nonlocal properties (in both space and time).  We emphasize that, we believe, these can only be measured using the techniques of weak measurements introduced here.

Finally, the second method concerning analysis of changes in probability distributions  discussed in \S II.b (i.e. using  the Fourier transform of the conserved quantity) has many advantages when compared to the traditional analysis with respect to moments discussed in \S II.a. For example, the Fourier transform method 
provides us with the parameters relevant to physical problem (e.g. the distance $D$ between the slits), while the first ``moments" approach remains silent. 
(We note that, in effect, looking at the modular
   variables is asking how the Fourier transform of the momentum distribution changes.)  In addition, there are different conservation laws involved with the second method which are more relevant and useful.  One of the basic notions used in the analysis of conserved quantities in any interaction is that as the probability of one conserved quantity changes ($prob(A)$), then the probability of another should also change ($prob(B)$), such that the probability of the sum ($prob(A+B)$) does not change.  
As we pointed out, there are situations where the probability of one variable does not change ($prob(B)$), while the probability of the other does change ($prob(A)$).  This, per Theorem II, can only happen if that variable (e.g. $B$) is completely uncertain.  
That is, this can only happen if the fourier transform of $prob(B)$ below some value remains unaffected while the fourier transform of  $prob(B)$ above some value is affected.  
This means that there is a whole range of modular variables that are being exchanged nonlocally and a large number of conservation laws which can be utilized.

\section{\bf VI. Conclusion}

The purpose of this short, yet self-contained note, is to highlight some new aspects of interference using pre- and post-selection and weak measurements.  
In particular, we calculated that, if the left slit is later closed, then the probability that the earlier weak measurement shifts any particles from the right slit to the left slit is $O(\frac{1}{N})$.  Therefore, in the limit of large $N$, the weak measurement does not shift even a single particle from the right slit to the left slit.  This can be confirmed by placing a photographic plate at the left slit.
On the other hand, if the left slit is later opened (i.e. after the weak measurement), then we calculated that 
 a small number of particles (independent of $N$), are shifted from the right slit to the left slit.  However, all $N$ particles contribute to the dramatically different weak measurement results.

Our use of the Heisenberg picture lead us to a {\bf physical} explanation for the different behaviors of a {\em single} particle when the distant slit is open or closed:  instead of having a quantum wave that passes through all slits, we have a localized particle with {\em non-local} interactions with the other slit(s).
Although  particles localized around the right  slit can exchange modular momentum (non-locally) with the ``barrier" at the left slit, the uncertainty in quantum mechanics appears to be just right to protect causality.

{ While the Heisenberg picture and the Schr\"{o}dinger pictures are equivalent formulations of quantum mechanics, nevertheless, the results discussed here support a new approach which 
has led to new insights, new intuitions, new experiments, and even the possibility of new devices that were missed from the old perspective. These types of developments are signatures of a successful re-formulation.

\bigskip

\noindent We thank J. Gray, A. D. Parks, S. Spence, and J. Troupe.

{
\appendix

\bigskip

\addcontentsline{toc}{section}{{{\bf \large Appendices}}}

\bigskip

\noindent {\bf APPENDICES}

\bigskip

\addtocounter{section}{1}


\noindent{\bf Appendix A:} a basic theorem which characterizes all interference phenomenon is that
all {\bf moments} of both position and momentum are independent of the relative phase 
   parameter $\alpha$.   It is easy to see this for the particular double-slit wavefunction eq. \ref{eq1} assuming  there is no overlap of $\psi_{\mathrm{L}}(x,0)$
   and $\psi_{\mathrm{R}}(x,0)$ and that $n$ is an integer, then for all values of $t$, and choices of $\alpha\,,\,\beta$:
   \begin{equation}\label{9.6}
      \int[\Psi_\alpha^*(x,t)\Psi_\alpha(x,t)-\Psi_\beta^*(x,t)\Psi_\beta(x,t)]x^ndx=0
\label{thminterference}
   \end{equation}
\label{proofinterference}
We see that $\Psi_\alpha^*(x,0)\Psi_\alpha(x,0)$ is
   independent of $\alpha$, and hence
   $\Psi_\alpha^*(x,0)\Psi_\alpha(x,0)-\Psi_\beta^*(x,0)\Psi_\beta(x,0)=0$. Therefore, at $t=0$,
   $\langle x^n\rangle$ is independent of $\alpha$, as is $\langle p^n\rangle$. The latter follows
   from the non-overlapping nature of 
$\psi_{\mathrm{L}}(x,0)$
   and $\psi_{\mathrm{R}}(x,0)$. It is also easy to show that
   $\langle x^np^m\rangle$ at $t=0$ is independent of $\alpha$ by using the Heisenberg
   representation
   $\langle x^n(t)\rangle=\int\Psi_\alpha^*(x,0)x^n(t)\Psi_\alpha(x,0)dx$
and noting that $x(t)=x(0)+p(0)\frac{t}{m}$ and $p(t)=p(0)$
   in this representation, we must have
      $\langle x^n(t)\rangle=\int\Psi_\alpha^*(x,0)[x(0)+p(0)\frac{t}{m}]^n(t)\Psi_\alpha(x,0)dx$.
   This is clearly independent of $\alpha$, since term by term it is independent of $\alpha$.
   Eq. (\ref{9.6}) then follows, 
and holds for $p^n$,
   as long as we retain the proper $\Psi_\alpha^*p^n\Psi_\alpha$ order.

\bigskip
\addtocounter{section}{1}

\noindent{\bf Appendix B:} 
 The ``configuration" space states are those of a
particle at N discrete locations: $|\Psi(1)\rangle$ at slit 1, $|\Psi(2)\rangle $ at slit
two, etc., and the $N$ discrete modular momentum eigenstates are the appropriate
linear combinations:
\begin{eqnarray} 
\!\!\!\!\!\!\!|\chi(1)\rangle &=& \,\,\frac{1}{\sqrt{N}} [|\Psi(1)\rangle + |\Psi(2)\rangle +\cdots+|\Psi(N)\rangle ]\!\!\!\!\!\!\!\\ 
|\chi(2)\rangle &=&\frac{1}{\sqrt{N}} [|\Psi(1)\rangle + b|\Psi(2)\rangle +\cdots+ b^{N-1} |\Psi(N)\rangle]\nonumber\\ 
|\chi(3)\rangle &=&\frac{1}{\sqrt{N}} [|\Psi(1)\rangle +b^2|\Psi(2)\rangle +\cdots b^{2(N-1)}|\Psi(N)\rangle ] \nonumber\\
 &\vdots& \nonumber\\ 
|\chi(n+1)\rangle &=&\frac{1}{\sqrt{N}} [|\Psi(1)\rangle +\cdots b^{n(N-1)}|\Psi(N)\rangle ], \nonumber\\
 &\vdots& \nonumber
\end{eqnarray} 
where $b=\exp(-i2\pi/N)$, and each of the  $|\chi(k)\rangle$ being an eigenstate of the cyclic shift
operator $1\rightarrow 2\rightarrow 3\rightarrow ...\rightarrow N\rightarrow 1$ namely the relevant modular operator
with eigenvalues $b^{k-1}$.
 The inverse of the above, relates each of the configuration eigenstates
to an {\em equal} weight combination of the $|\chi(k)\rangle$ states which again is a
state with maximal angular momentum uncertainty.
 (This in turn is analog of the Dirac $\delta(x)$ function being an equal
 weight superposition of all regular continuous momentum $p$, for the
discrete Kronecker $\delta(j)$ in the present case.)

\bigskip

\addtocounter{section}{1}

\noindent{\bf Appendix C: The non-locality of modular variables is generic to all interference phenomenon}

{
Although much of the discussion in this article focuses on the simplest interference example with 2-slits, our approach becomes clearer when it is applied to an infinite number of slits with (interfering) particles that are initially in an eigenstate of momentum.  
In this case, we can directly speak about modular momentum (instead of slightly more complicated functions for the double-slit setup).  Also, 
both the non-local equation of motion for modular momentum is exact as is the conservation of modular momentum.

Consider a system of infinite slits~\cite{danny} that is freely moving in the x-direction (see fig. \ref{modmom2}.  Suppose particles are sent towards the slits in a momentum eigenstate $p_y\equiv p_o$ and therefore $p_x=0$.  
Once again, we see that superposing any countable number of wavepackets (so that $\Delta x$ becomes arbitrarily large) need not change the uncertainty in momentum or any moment of momentum.
After passing through the slits, one can prove that the transverse momentum of the particles is $p_x=n\frac{h}{D}$.
Therefore, the particle and slit-grating can only exchange transverse momentum in integer multiples of $\frac{h}{D}$.  
With a Hamiltonian $H=\frac{p^2}{2m}+ V(x)$ (where $V(x+D)=V(x)$), one can also prove that $[H,e^{\frac{i}{\hbar}\hat{p}D}]=0$, i.e. modular momentum is conserved.

As we mentioned earlier, we can open or close other slits, or in close analogy, perform some operation (e.g. applying an uncertain potential).  It is easier to see the non-locality of modular variables if we perform the later, as in an Aharonov-Bohm setup.  Suppose then that we now place solenoids with magnetic flux $\Phi=\frac{1}{2}\Phi_o$ inside the slit-gratings (see fig. X) so that there is no contact between the particles and the solenoids or their fields.  
Suppose that we connect all the solenoids together so that they could move independent of the slit-gratings.
One can prove that the condition for constructive interference is satisfied if the transverse momentum exchange is $\frac{h}{D}\left\{n+\frac{\Phi}{\Phi_o} \right\}=\frac{h}{D}\left\{n+\frac{1}{2} \right\}$.  We know that without the solenoids, only an integer multiple of $\frac{h}{D}$ was exchanged between the particles and the slit-grating.  This means that $\frac{h}{2D}$ is exchanged non-locally between the solenoids and the particles due to conservation of modular momentum.

We also know that, by definition, $\Delta$ (position of the solenoids)$\leq D$ and therefore $\Delta$ (momentum of solenoids)$\geq \frac{h}{D}$.  If we send a single electron, then $\frac{h}{2D}$ is exchanged non-locally with the solenoids.  This is less than the uncertainty $\frac{h}{D}$ and is therefore un-observable.  However, if we send many electrons,
then would the exchange become observable, as happens, e.g. in $\sqrt{N}$ average displacement in a random walk?
Again, this might happen if ordinary momentum were involved, but modular momentum is constrained.

\vskip-.1cm
\begin{figure}[tbph] 
  \centering
\includegraphics[width=\columnwidth]{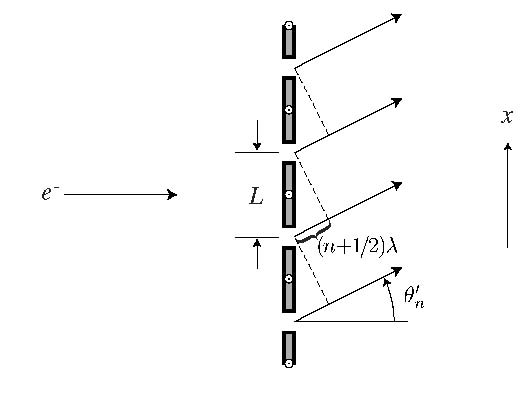}
\caption[]{\footnotesize }  
\label{modmom2}
\vskip-.5cm
\end{figure}

When the electron passes through the slit-gratings then we do not know through which slit it passed.  We only know its position modulus the distance $D$ between the slits.  We can define modular position as $x_{\mathrm{mod}}\equiv x$ modulus $D$. Ordinary position then can be expressed as: $x=N_x D + x_{\mathrm{mod}}$.
The grating is providing us with information about both $x$ mod $D$ and also $p$ mod $\frac{h}{D}$.  

To see how modular variables accomplish this provides further motivation for the significance of modular variables and also answers the question that began this section ``which operators become uncertain when WWM information is obtained".  Suppose we seek those functions $f(p)$ and $g(x)$ such that $[f(p),g(x)]=0$.  If the functions commute, then we could obtain information about $f(p)$, and, at the same time, $g(x)$.  Classically there is no non-trivial answer.  However, using the quantum commutator, we see that if $f(p)=f(p+p_o)$ and $g(x)=g(x+x_o)$ and if $x_op_o=h$, then $f(p)$ commutes with $g(x)$.  We therefore see that these periodic functions lead us to the notion of modular variables.

We may thus derive the following relations:
\beq
[e^{\frac{i}{\hbar}p_{\mathrm{mod}}D},N_x]=e^{\frac{i}{\hbar}p_{\mathrm{mod}}D}
\label{emod2}
\eeq
and
\beq
[e^{i2\pi x_{mod}/D},N_p]=e^{i2\pi x_{mod}/D}
\eeq
These two commutation relations are similar to the angular momentum-angle relation $[L_z,e^{i\phi}]=e^{i\phi}$: if the angular momentum is known precisely then the angle is completely uncertain.  Thus, $p_{\mathrm{mod}}$ is conjugate to $N_x$ and $x_{mod}$ is conjugate to $N_p$:
\beq
\Delta N_x \Delta p_{\mathrm{mod}}\geq \frac{h}{D}
\label{dntdem2}
\eeq
Which means, e.g. that if the system is localized in space, i.e. $N_x$ is known, then the modular momentum is completely uncertain.  Another uncertainty relation that can be derived is:
\beq
\Delta N_p \Delta x_{mod}\geq D
\eeq
I.e. if the momentum is well known, then the modular position is uncertain.

The standard Heisenberg uncertainty principle does not allow us to localize a particle in phase space to anything less than an area of $\hbar$.  The uncertainty principle for modular variables allows us to precisely locate a spot within a cell of area $\hbar$, i.e. $x_{mod}$ and $p_{\mathrm{mod}}$ are certain, but it gives no information about which cell it is in, i.e. $N_x$ and $N_p$ are completely uncertain.  This suggests that we can have precise but partial information about the momentum, namely the modular momentum, and simultaneously precise but partial information about the location, namely the modular position (see figure \ref{nlphasespace}.a).   We can also have intermediate situations, i.e. less precise information of the exact point within a phase cell but more information of which cell we're in (see figure \ref{nlphasespace}.b).

\begin{figure}[tbph] 
  \centering
\includegraphics[width=\columnwidth]{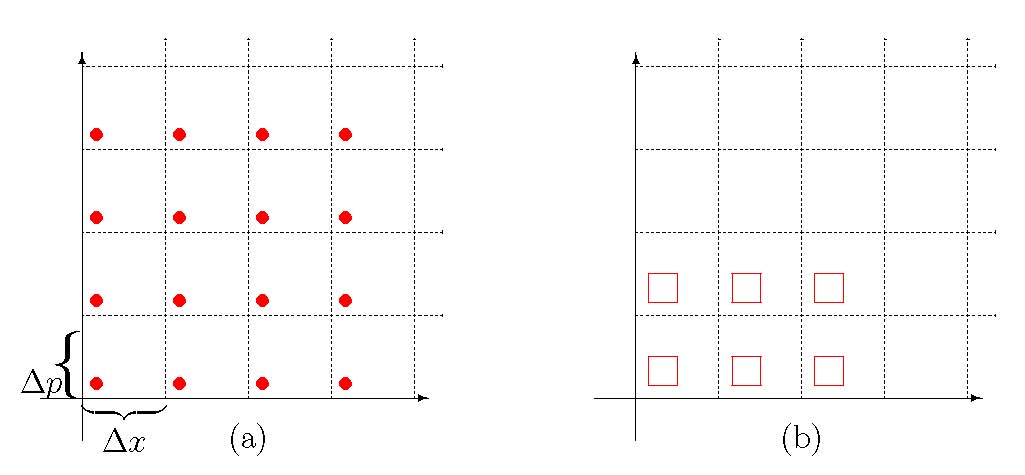}
\caption[Definiteness of modular variables in phase space]{\small (a) phase space of particles with definite modular position, $x$ mod $D=D/4$ and definite modular momentum $p$ mod $h/D=\frac{1}{2}h/D$ and cell size $\Delta x\Delta p=\hbar$;  (b) intermediate situation: more knowledge of which cell we are in means  less knowledge of modular variable}
\label{nlphasespace}
\end{figure}

With single wavepackets, $N_p$ and $N_x$ are well known and therefore $x_{\mathrm{mod}}$ and $p_{\mathrm{mod}}$ are almost completely uncertain.  However, as these uncertainty relations indicate, when we have wavefunctions with more than one ``lump," then we must use $x_{\mathrm{mod}}$ and $p_{\mathrm{mod}}$.  In particular, consider again the example of infinite slits: the electron passes the grating, so we have precise information about $x_{\mathrm{mod}}$ (in
the transverse direction); $x_{\mathrm{mod}}$ and $p_{\mathrm{mod}}$ commute, so we have precise information about $p_{\mathrm{mod}}$.
But $x_{\mathrm{mod}}$ and $N_p$ do not commute (see eqs. X) so we have no information about $N_p$.
The interaction of the electron with the grating conserves $p_{\mathrm{mod}}$  so these facts
determine the interference pattern completely: $p_{\mathrm{mod}}$ fixes the position of the fringes relative
to the grating; $N_p$ is completely uncertain and therefore the fringes are equally dense. Now
consider the effect of a lattice of solenoids. The solenoids affect the modular momentum in the
same way as the stair potential $V(x)$ of fig. X. The nonlocal interaction of the electrons with
the solenoids changes $p_{\mathrm{mod}}$ of the diffracting electrons; hence the diffraction pattern shifts.

\addtocounter{section}{1}

\noindent{\bf Appendix D: Conservation law for modular variables}

Modular variables have different kinds of conservation laws that are enforced by the non-local equations of motion, and this will prove to be crucially important for this article.  
For the 2-slit setup, conservation of modular momentum is particular easy.  If we start with 
$\ket{\psi_{\mathrm{L}}} +\ket{\psi_{\mathrm{R}}}$, then we'll end up with $\ket{\psi_{\mathrm{L}}} +\ket{\psi_{\mathrm{R}}}$.  If we start with 
$\ket{\psi_{\mathrm{L}}} -\ket{\psi_{\mathrm{R}}}$, then we'll end up with $\ket{\psi_{\mathrm{L}}} -\ket{\psi_{\mathrm{R}}}$
More generally, the modular momentum analogy 
to conservation of ordinary 
 momentum (e.g. $P^{\mathrm{in}}_1+P^{\mathrm{in}}_2=P^{\mathrm{fin}}_1+P^{\mathrm{fin}}_2$) can be derived as follows. Using $\pi_1(P_1)=\cos (2\pi P_1/P_0)$ and $\pi_2(P_2)=\cos (2\pi P_2/P_0)$ (another expression for $p_{\mathrm{mod}}$) we see that:
\beq
\cos\lbrack 2\pi\lbrace P_1+P_2\rbrace/P_0\rbrack=\cos\lbrack 2\pi\lbrace P_1'+P_2'\rbrace/P_0\rbrack
\eeq
in other words:
\beq
\pi_1\pi_2-\sqrt{1-\pi_1^2}\sqrt{1-\pi_2^2}=\pi_1'\pi_2'-\sqrt{1-(\pi_1')^2}\sqrt{1-(\pi_2')^2}
\eeq
which gives:
\begin{eqnarray}
&&(\pi_1')^2+(\pi_2')^2-2\cos[ 2\pi\{ P_1+P_2\}/P_0]\pi_1'\pi_2'=\nonumber\\
&=&1-\cos[ 2\pi\{P_1+P_2\}/P_0]^2
\end{eqnarray}
Thus instead of a line $P_1+P_2=constant$, the conservation law for modular variables is an ellipse (see figure \ref{ellipse}).}  
\begin{figure}[tbph] 
  \centering
\includegraphics[width=\columnwidth]{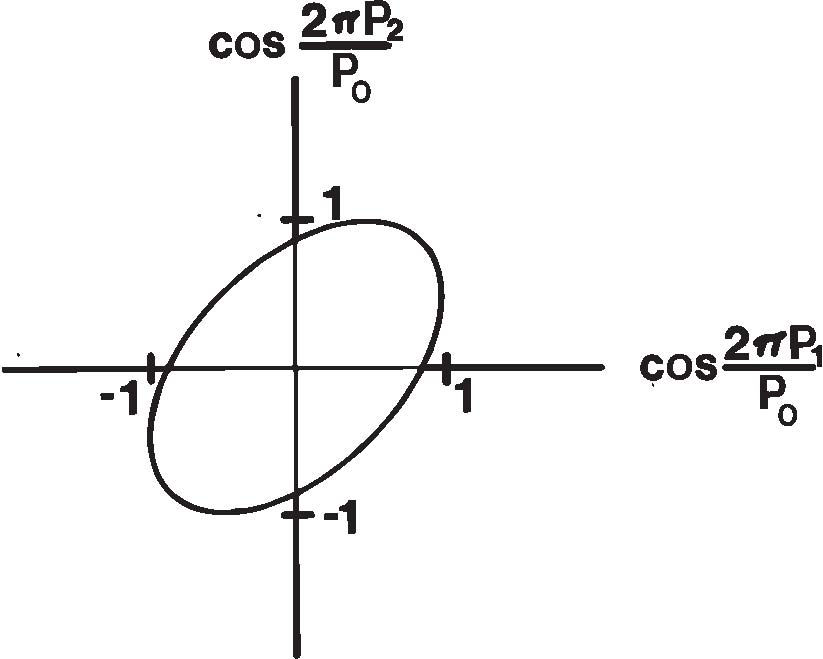}
\caption{Conservation Law for Modular Momentum}
\label{ellipse} 
\end{figure}

\vskip-.1cm

If we know the initial values of the modular
momentum of the two interacting systems, then we
may represent their initial state by a point on the
conserved ellipse of fig. \ref{ellipse}. As the interaction
between the two systems proceeds, the point
representing the system will move along the
ellipse and eventually come back to its original
position. We see then how the periodicity of
the non-local phenomena is reflected in the
conservation laws for the relevant modular
variables. We also note that in the classical 
limit $p_o\rightarrow 0$, so that $p$ mod $p_o$ changes so
rapidly as a function of $p$ as to become entirely
unobservable.

}


\begin{thebibliography}{99}
\bibitem{omit3} Unitarity  requires a reflected state $\ket{\psi_3}$. Unlike $\ket{\psi_{\mathrm{L}}}$ and $\ket{\psi_{\mathrm{R}}}$, moving in the $+y$ direction, $\ket{\psi_3}$ moves in the $-y$ direction.  Since $\bra{\psi_3}\psi_{\mathrm{L}}\rangle=0$, omitting it does not affect the argument.
\bibitem{qis} Los Alamos Quantum Computing Road map\cite{doolen}: ``At least two important precursors to this [quantum computing] paradigm shift had critical influence," citing non-locality, e.g. the Einstein-Podolsky-Rosen/Bohm~\cite{shimony} (EPRB) and Aharonov-Bohm (AB) effects, and developments in quantum information theory.
\bibitem{ya-heis} {Yakir Aharonov described this general approach to interference 
to Heisenberg personally.  Prof. Heisenberg had never thought before about interference phenomenon in the Heisenberg picture and was extremely pleased by it.}
\end{thebibliography}
\end{document}